\documentclass[twocolumn, amsmath,amssymb, superscriptaddress]{revtex4}

\usepackage{dcolumn}% Align table columns on decimal point
\usepackage{bm}% bold math
\usepackage{amsmath}
\usepackage{amsfonts}
\usepackage{graphics}
\usepackage[dvips]{graphicx}
\usepackage{times}
\usepackage{setspace}
\usepackage{color}   % {\color{red} TEXT }
\usepackage{verbatim}
\usepackage[raggedright]{titlesec}
\usepackage[normalem]{ulem}
\usepackage{framed}

\usepackage[breaklinks=true]{hyperref}
\usepackage{breakurl}

\usepackage{balance}

\hypersetup{colorlinks=true,citecolor=blue, linkcolor=black,urlcolor=blue}

\begin{document}
%\title{The Apostle Effect: Quantifying the impact of super ties in scientific careers}
\title{Quantifying the impact of weak, strong, and super ties in scientific careers}
\author{Alexander M. Petersen}
\affiliation{Laboratory for the Analysis of Complex Economic Systems, IMT Lucca Institute for Advanced Studies, Lucca 55100, Italy}

\begin{abstract} 
Scientists are frequently faced with the important decision to start or terminate a creative partnership. This process can be influenced by strategic motivations, as early career researchers are pursuers, whereas senior researchers are typically  attractors, of new collaborative opportunities. Focusing on the longitudinal aspects  of scientific collaboration, we analyzed 473 collaboration profiles using an ego-centric perspective which accounts for researcher-specific characteristics and  provides insight into a range of topics, from career achievement and sustainability to team dynamics and efficiency.  From more than 166,000 collaboration records, we quantify the frequency distributions of collaboration duration and tie-strength, showing that collaboration networks are dominated by weak ties characterized by high turnover rates. We use analytic extreme-value thresholds to identify a new class of indispensable `super ties', the strongest of which commonly exhibit $>50$\% publication overlap with the central scientist. The prevalence of  super ties  suggests that they arise from career strategies based upon cost, risk, and reward sharing and complementary skill matching. We then use a combination of descriptive and  panel regression methods to compare the subset of publications coauthored with a super tie to the subset  without  one, controlling for pertinent features such as career age, prestige, team size, and prior group experience. We find that  super ties contribute to above-average productivity and a 17\% citation increase per publication, thus identifying these partnerships --  the analog of life partners -- as a major factor in science career development.
\end{abstract}
%\date{\today}

%Currently 233
%Sci Reports 150 word abstract: 
%The Methods section is limited to 1500 words. Figure legends are limited to 350 words. References are limited to 60. Footnotes are not used.
%Depending on the word count, Articles may have up to 8 display items (figures and/or tables).
%PNAS: 250 word abstract, 10 pages for PNAS Plus
%NatureComm: 150 word abstract
%The main text (not including abstract, Methods, References and figure legends) is limited to 5,000 words. The maximum title length is 15 words. The abstract Ñ which should be no more than 150 words long and contain no references Ñ should serve both as a general introduction to the topic and as a brief, non-technical summary of the main results and their implications.
%The main text of an Article should begin with an introduction (without heading) of referenced text that expands on the background of the work (some overlap with the abstract is acceptable), followed by sections headed Results, Discussion (if appropriate) and Methods (if appropriate). The Results and Methods sections may be divided by topical subheadings; the Discussion should be succinct and may not contain subheadings. Methods are typically less than 3000 words. Figure legends are limited to 350 words each. References are limited to 70. Footnotes are not used.
%Science Online Only Research Article: These can be longer, up to 8000 words and include methods and additional figures as part of the main presentation. The cover letter should indicate why the additional length is merited.

\maketitle

%\footnotetext[1]{ Send correspondence to:  \text{petersen.xander@gmail.com}}

\begin{framed} 
\noindent {\bf A scientist will encounter many potential collaborators throughout the career. As such, the choice to start or terminate a collaboration can be an important strategic consideration with long-term implications. While previous studies  have focused primarily on aggregate cross-sectional collaboration patterns, here we analyze the collaboration network  from a researcher's local perspective along his/her career. Our  longitudinal approach reveals that scientific collaboration is characterized by a high  turnover rate  juxtaposed with surprisingly frequent  `life partners'.  We  show that these extremely strong  collaborations have a significant positive impact on   productivity and citations -- the apostle effect -- representing the advantage of `super' social ties characterized by trust, conviction, and commitment.} 
For the Supporting Information see the published version: A. M. Petersen (2015) {\it Proc. Nat. Acad. Sci. USA} 112, E4671--E4680. %{\bf 3}: 24. 
\href{http://www.pnas.org/content/112/34/E4671.abstract}{DOI:10.1073/pnas.1501444112}
%\text{Author contact: petersen.xander@gmail.com}
\end{framed}
%\end{comment}

Science operates at multiple  scales, ranging from the global and institutional scale  down to  the level of groups and individuals \cite{borner_multi-level_2010}.  
%At the top level of organization, science policy is a major driving force, acting over long time scales and distances.  
%As one moves to the micro scale, social interactions -- between the principal-agents of large scale organizations  (funding bodies, universities, publishers, etc.) and the  large population of researchers who are  connected within the `invisible college'  -- represent a fundamental impetus for the evolution of science.
Integrating this system are multi-scale social networks that are ripe with structural, social, economic, and  behavioral complexity \cite{stephan_how_2012}.
A subset of this multiplex is the scientific collaboration network, which forms the structural foundation for social capital investment, knowledge diffusion,  reputation signaling, and important mentoring relations  \cite{nahapiet_social_1998,wuchty_increasing_2007,petersen_reputation_2014,malmgren_role_2010,petersen_quantitative_2014,pavlidis_together_2014}. 

%The social ties comprising these networks  can be  categorized according to 4  typologies: similarities, social relations, interactions, and flows  \cite{borgatti_network_2009}. % outlined by Borgatti et al.
Here we focus on collaborative endeavors that result in scientific publication, a process which  draws on various aspects of social ties, e.g. colocation, disciplinary identity, competition, mentoring, and knowledge flow \cite{borgatti_network_2009}. 
The dichotomy between  strong  and weak  ties is a longstanding  point of research \cite{granovetter_strength_1973}. However, in `science of science' research, most  studies have analyzed macroscopic collaboration networks aggregated across time, discipline, and individuals  \cite{newman_structure_2001,newman_scientific_2001-3, Barabasi_evolution_2002,newman_coauthorship_2004,guimera_team_2005,palla_quantifying_2007,pan_strength_2012,martin_coauthorship_2013,ke_tie_2014,borner_simultaneous_2004,milojevic_principles_2014}.  Hence, despite these significant efforts, we know little about  how properties of the local social network  affect scientists'  strategic career decisions. For example, how might  creative opportunities in the local collaboration network impact a researcher's decision to explore new avenues versus exploiting old partnerships, and what may be the career tradeoffs in the short versus the long-term, especially considering that academia  is driven by dynamic knowledge frontiers \cite{march_exploration_1991,lazer_network_2007}.
 
Against this background,  we develop a  quantitative approach for improving our understanding of the role of weak and strong ties, meanwhile uncovering a third  classification -- the `super tie' -- which we find to  occur rather frequently. We analyzed longitudinal career data for researchers from  cell biology and physics, together comprising a set of 473 researcher profiles spanning  more than 15,000 career years, 94,000 publications, and 166,000 collaborators. In order to account for prestige effects, we define 2 groups within each discipline set, facilitating a comparison of  top-cited scientists with  scientists that are more representative of the entire researcher population (henceforth referred to as ``other''). From the $N_{i}$ publication records spanning the first $T_{i}$ career years of each central scientists $i$, we constructed longitudinal  representations of each scientist's coauthorship history. 

We adopt an ego-centric perspective in order to track research careers  from their inception along their longitudinal growth trajectory.
By using a  local perspective we  control for the heterogeneity in collaboration patterns that exists  both  between and within disciplines. We also control  for other career-specific  collaboration and productivity differences that would otherwise be averaged out by aggregate cross-sectional methods. 
 Thus, by simultaneously  leveraging multiple features of the data -- resolved over the dimensions of time, individuals,  productivity, and citation impact -- our  analysis   contributes to the literature on science careers as well as   team activities characterized by dynamic   entry and exit of human, social, and creative capital. Given that collaborations in business, industry, and academia are increasingly operationalized via team structures, our findings provide relevant quantitative insights into the mechanisms of team formation \cite{guimera_team_2005},  efficiency \cite{petersen_persistence_2012}, and  performance \cite{pentland_new_2012,woolley_evidence_2010}. 

%Understanding how dynamic collaboration patterns are elated to variables such as  collaboration size, the level of hierarchical management, and other individual-level idiosyncrasies such as leadership style, remains a difficult challenge.  
%CITE THESE NEW \cite{schulz_exploiting_2014,sarigl_predicting_2014,rivera_dynamics_2010,nahapiet_social_1998 , shane_network_2002 XX,  conduit for exploring new ventures \cite{shane_network_2002}} newman_clustering_2001 XX individual profile analysis XX \cite{ductor_social_2014} using collaboration network measures to improve predictive power of future productivity (although results are weak)  dynamic collaboration patterns and hierarchical management
 The organization of our study is structured as follows.
The longitudinal nature of a career requires that we start by  quantifying the  tie-strength between two collaborators from two different perspectives: duration and strength. First we analyze the collaboration duration, $L_{ij}$, defined as the time period between the first and last publication between  two researchers $i$ and $j$. Our results indicate that the  ``invisible college'' defined by collaborative research activities (i.e. excluding informal communication channels and  arm's length associations)  is surprisingly dominated by high-frequency interactions lasting only a few years.  
We then focus our analysis on the collaborative `tie strength', $K_{ij}$, defined as the cumulative number of publications coauthored by $i$ and $j$ during the  $L_{ij}$ years of activity. 

 From the entire set of collaborators, we then  identify  a subset of  `super tie'  coauthors -- those $j$  with $K_{ij}$ values that are statistically unlikely   according to an author-specific  extreme-value criteria.  
 %Following from the analytic features of the distribution of  $K_{ij}$ values within each profile $i$,  
 %In other words, the super ties are defined according to author-specific  extreme-value criteria,  
 %This approach provides an objective means to  identify, and thus quantify, the role and impact of super ties on scientists' careers.
 Because almost all of the researchers we analyzed have more than one super tie, and roughly half of the publications we analyzed include at least one super-tie coauthor, we were able to quantify the added value of super  ties --   for both for productivity and citation impact --   in two ways, (i) using descriptive measures and (ii) implementing a fixed-effects regression model.
  Controlling for author-specific features, we find that  super ties are associated with  increased publication rates and increased citation rates.

  We term this  finding the  `apostle effect',  signifying  the  dividends generated by  extreme social ties based upon  mutual trust, conviction, and commitment.  
  This term borrows from  biblical context, where an apostle represents a distinguished partner selected  according to his/her noteworthy attributes from among a large pool of candidates.  What we do not connote is any  particular power relation  (hierarchy)  between $i$ and the super tie coauthors, which is beyond the scope of this study. Also, because the perspective is centered around  $i$, our super-tie definition is not symmetric, i.e. if $j$ is a super tie of $i$, $i$ is not necessarily a super tie of $j$.
%; here it is ememphasized that such  represent a social tie with noteworthy attributes  that are

Because   super ties   have  significant long-term impact on productivity and citations,  our results are important  from a career development perspective, reflecting the strategic benefits of cost,  risk, and reward-sharing via long-term  partnership. The implications of  research partnerships  will become increasingly relevant   as more careers become inextricably embedded in team science environments, wherein it can be difficult to identify contributions, signal achievement, and distribute credit.
The  credit distribution problem has received recent attention from the perspectives of  institutional policy \cite{pavlidis_together_2014}, team ethics \cite{petersen_quantitative_2014}, and practical  implementation  \cite{stallings_determining_2013,allen_credit_2014,shen_collective_2014}.

\section*{Results}
\noindent {\bf Defining the ego collaboration network.}  
%Most models of scientific collaboration focus on the aggregate scale \cite{newman_structure_2001,newman_scientific_2001-3, newman_coauthorship_2004,palla_quantifying_2007,pan_strength_2012, martin_coauthorship_2013,ke_tie_2014,borner_simultaneous_2004,milojevic_principles_2014}.
Our framework assumes the perspective of the central scientist  $i$ in the ego network formed by all of his/her collaborators (indexed by $j$). 
We use  longitudinal publication data from Thompson Reuters Web of Knowledge (TRWOK),   comprising 193 biology and 280 physics careers.  Each career profile is constructed by aggregating the collaboration metadata  over the first  $t=1 \dots T_{i}$ years of his/her career. 
 We  downloaded the TRWOK data in calendar year $Y_{i}$, which is the citation count census year. Each disciplinary set includes a subset of  100 highly-cited scientists (hereafter referred to as ``top''), selected using a ranking of the top-cited researchers in the high-impact journals {\it Physical Review Letters} and {\it Cell}. 
The rest of the researcher profiles (``other'') are aggregated  across physics and cell biology, with  subsets that are specifically active in the domains of graphene, neuroscience, molecular biology, and genomics.
The ``other'' dataset only includes $i$ with at least as many publications as the smallest $N_{i}$ among the top-cited researchers: as such,  $N_{i}\geq 52$ for biology and $N_{i}\geq 46$ for physics.
This facilitates a reasonable comparison between ``top'' and ``other'', possibly identifying differences attributable to innate success factors.  See  the Supporting Information Text  ({\it SI Text})  for further details on the data selection. 
%Comparisons between the ``top'' and ``other'' datasets will serve to indicate commonalities as well as differences, the latter possibly relating to innate success factors.

  \begin{figure}
\centering{\includegraphics[width=0.45\textwidth]{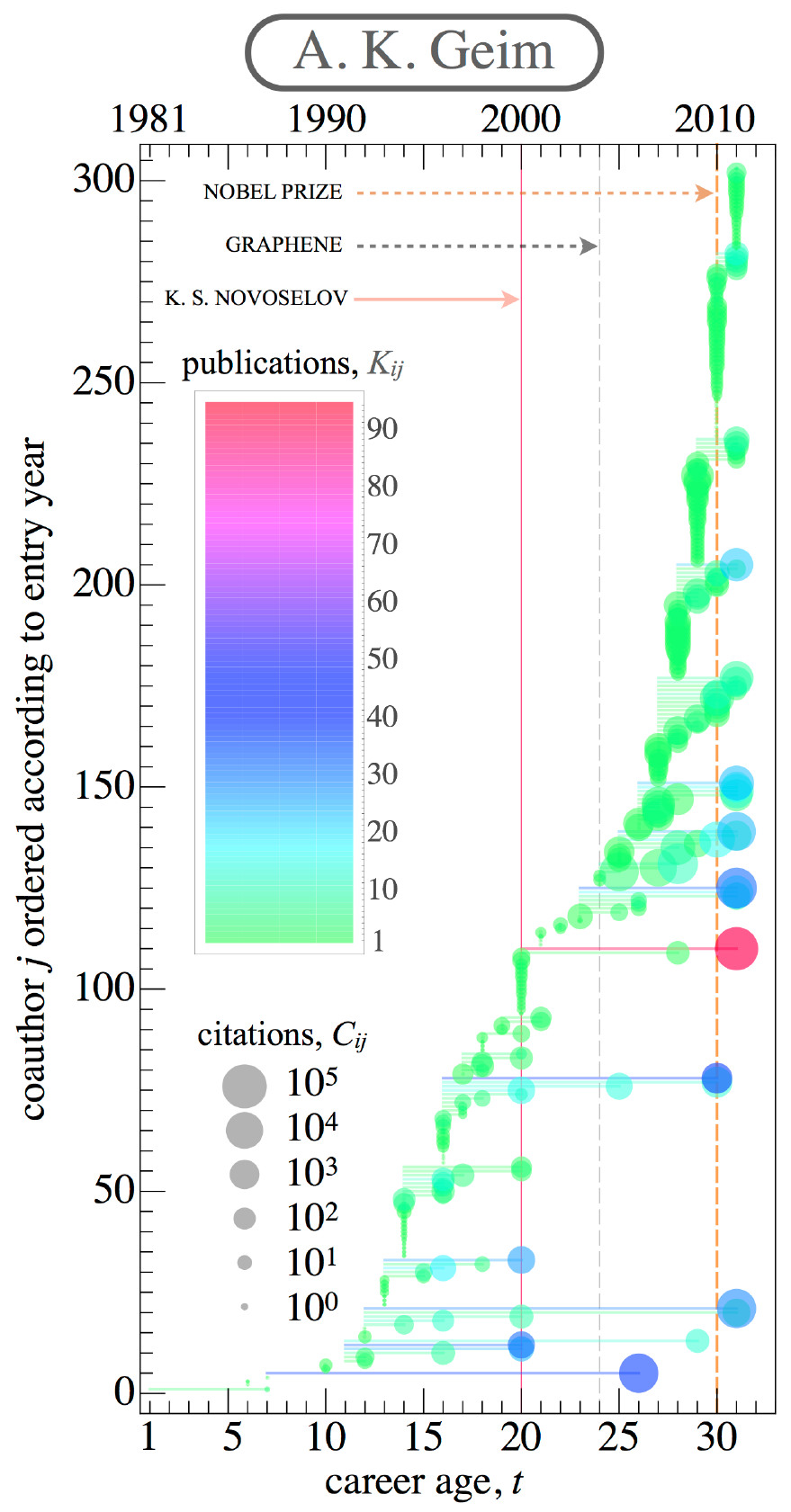}}
\caption{\label{GeimSchematic} {\bf Visualizing the embedding of academic careers in dynamic social networks.}  A career schematic showing A. Geim's collaborations, ordered by entry year. Notable career events include the first publication in 2000 with K. S. Novoselov (co-winner of the 2010 Nobel Prize in Physics) and their first graphene publication in 2004. %165  of 303 collaborations last  longer than the  2.1 year duration characteristic of Geim's collaborations, including for example, his spouse, I. V. Grigorieva since 1992. 
An interesting  network reorganization accompanies Geim's institutional move  from Radboud University Nijmegen (NL)  to U. Manchester (UK) in 2001. Moreover, the rapid accumulation of coauthors following the 2004 graphene discovery signals the new opportunities  that accompany reputation growth.
}
\end{figure}

This longitudinal approach leverages  author-specific factors,  revealing   how career paths are affected by idiosyncratic events. To motivate this point, Fig. \ref{GeimSchematic} illustrates the career trajectory of A. Geim, co-winner of the 2010 Nobel Prize in Physics. This schematic highlights three fundamental dimensions of collaboration ties --  duration, strength, and impact:  
\begin{enumerate}
\item[(a)] each horizontal line indicates the collaboration of length $L_{ij} \equiv t_{ij}^{f}-t_{ij}^{0}+1$ between  $i$ and coauthor $j$, beginning with their first joint publication
 in year $t_{ij}^{0}$ and ending with  their last observed joint publication in year
$t_{ij}^{f}$; 
\item[(b)] the circle color indicates the  total number of joint publications, $K_{ij}$,  representing our quantitative measure of `tie strength';
\item[(c)]  the circle size indicates the net citations $C_{ij}=\sum_{p} c_{j,p}$ in  $Y_{i}$, summed over all  publications $p$ that include $i$ and $j$.
\end{enumerate}
 Figs. S1 and S2 in the {\it SI Text} further illustrate the variability in collaboration strengths, both  between and within career profiles.
It is also worth mentioning that since multiple $j$ may contribute to the same $p$, it is possible for  coauthor measures to covary.
However, for the remainder of the analysis we focus on the dyadic relations between only $i$ and $j$, leaving the triadic and  higher-order `team' structures as an avenue for future work. For example, it would be interesting to know the likelihood of  triadic closure between any two super ties of $i$, signaling coordinated cooperation; or contrariwise,  low triadic closure rates  may indicate hierarchical organization around $i$.
\\

\begin{figure*}
\centering{\includegraphics[width=0.68\textwidth]{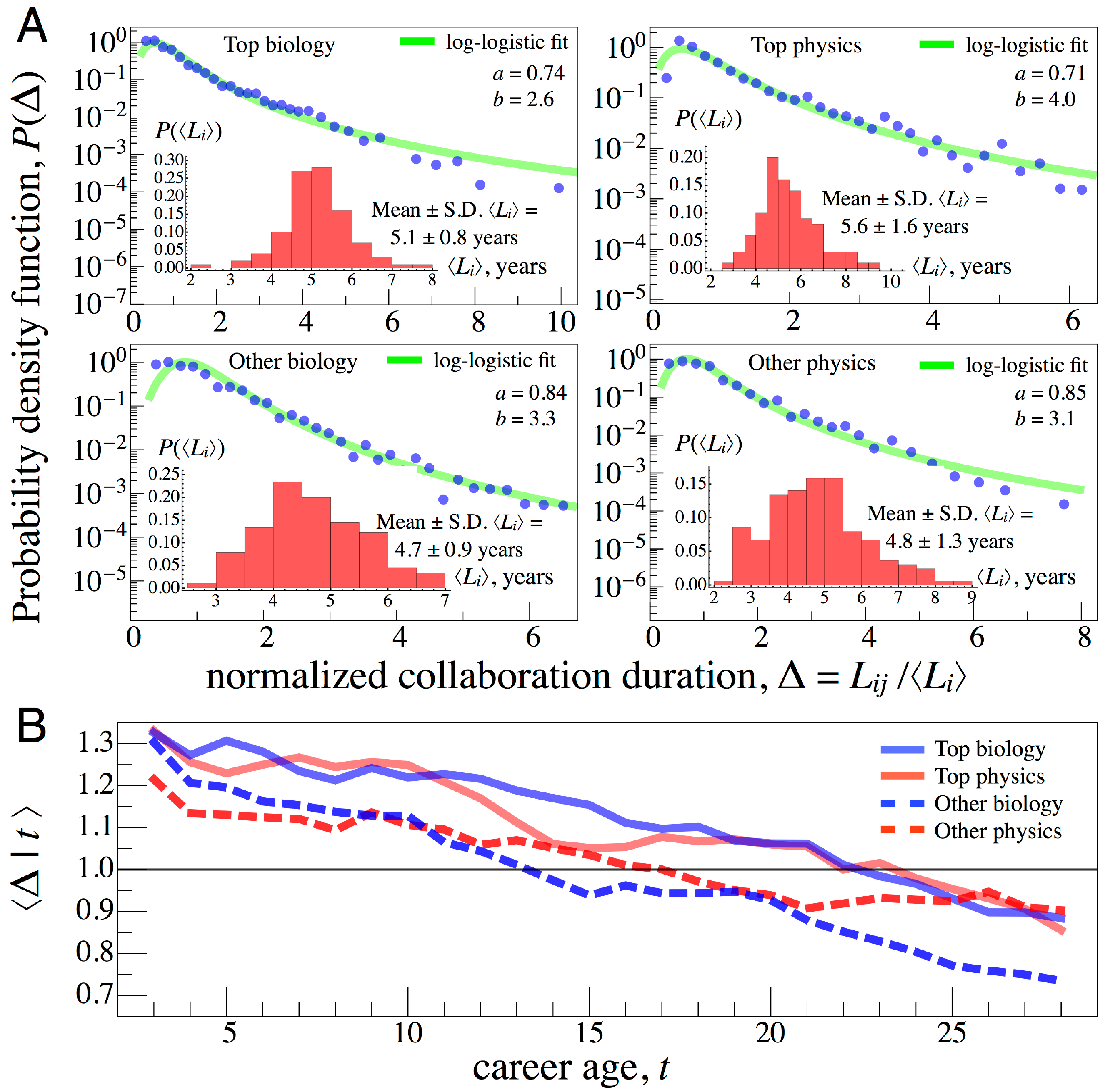}}
\caption{\label{LongPDF}  {\bf Log-logistic distribution of collaboration duration.}
  (A) The probability distribution $P(\Delta)$ is  right-skewed and well-fit by the log-logistic pdf defined in Eq.~[\ref{PDelta}]. (Insets) The probability distribution $P(\langle L_{i}\rangle)$ show that the characteristic collaboration length in physics and biology is typically between 2 and 6 years. 
  %Panel A created by trajectories from Longevity_AllData (without Lij1): 
(B) The decrease in the typical collaboration timescale, $\langle \Delta \vert t \rangle$, reflects how careers transition from being pursuers of collaboration opportunities to attractors of collaboration opportunities. 
}
\end{figure*}

\noindent{\bf Quantifying the  collaboration lifetime  distribution.} We use $L_{ij}$ to measure the duration of the productive interaction between $i$ and $j$. % over the first $T_{i}$ career years of $i$.
%The census date $Y_{i}$ corresponds to the date of the data download for profile $i$, and so the corresponding career year $T_{i}$ in some cases is past and in other cases is before career ending date. This detail is not important for the remainder of the analysis, since we are not analyzing the dynamics of the network, and so it is  not important to control for the career termination point (or phase).
%Also, in what follows, we use time-normalized citation measures and control for career age $t$ so that the author profiles are comparable and the descriptive statistics are aggregable. 
We find that a remarkable 60 to 80 percent of the collaborations have  $L_{ij}=1$ year (see {\it SI Text} Fig. S4).
Considering the overwhelming  dominance of the $L_{ij}=1$ events, in this subsection we concentrate our analysis  on the subset of repeat collaborations  ($L_{ij}>1$) which produced  two or more publications. 
Furthermore, due to censoring bias,   $L_{ij}$ values estimated for $j$ who are active  around the final  career year of the data ($T_{i}$) may be biased towards  small values.
 % (career profiles were downloaded in batches from 2010 to 2013. 
To account for this bias, in this subsection we also exclude  those collaborations that were active within the final $L^{c}_{i}$-year period, defining $L^{c}_{i}$ as an initial average $L_{ij}$ value calculated across all $j$ for each $i$. 
Then, we calculate a second representative mean value, $\langle L_{i} \rangle$, which is calculated excluding the  $j$ with $L_{ij}=1$ and the $j$  active in the final $L^{c}_{i}$-year  period.
Figure~\ref{LongPDF}(A) shows the probability distribution $P(\langle L_{i} \rangle)$, with  mean values ranging from 4 to 6 years, consistent with the typical duration of an early career   position (e.g. PhD or postdoctoral fellow, assistant professor). 

\begin{figure*}
\centering{\includegraphics[width=0.82\textwidth]{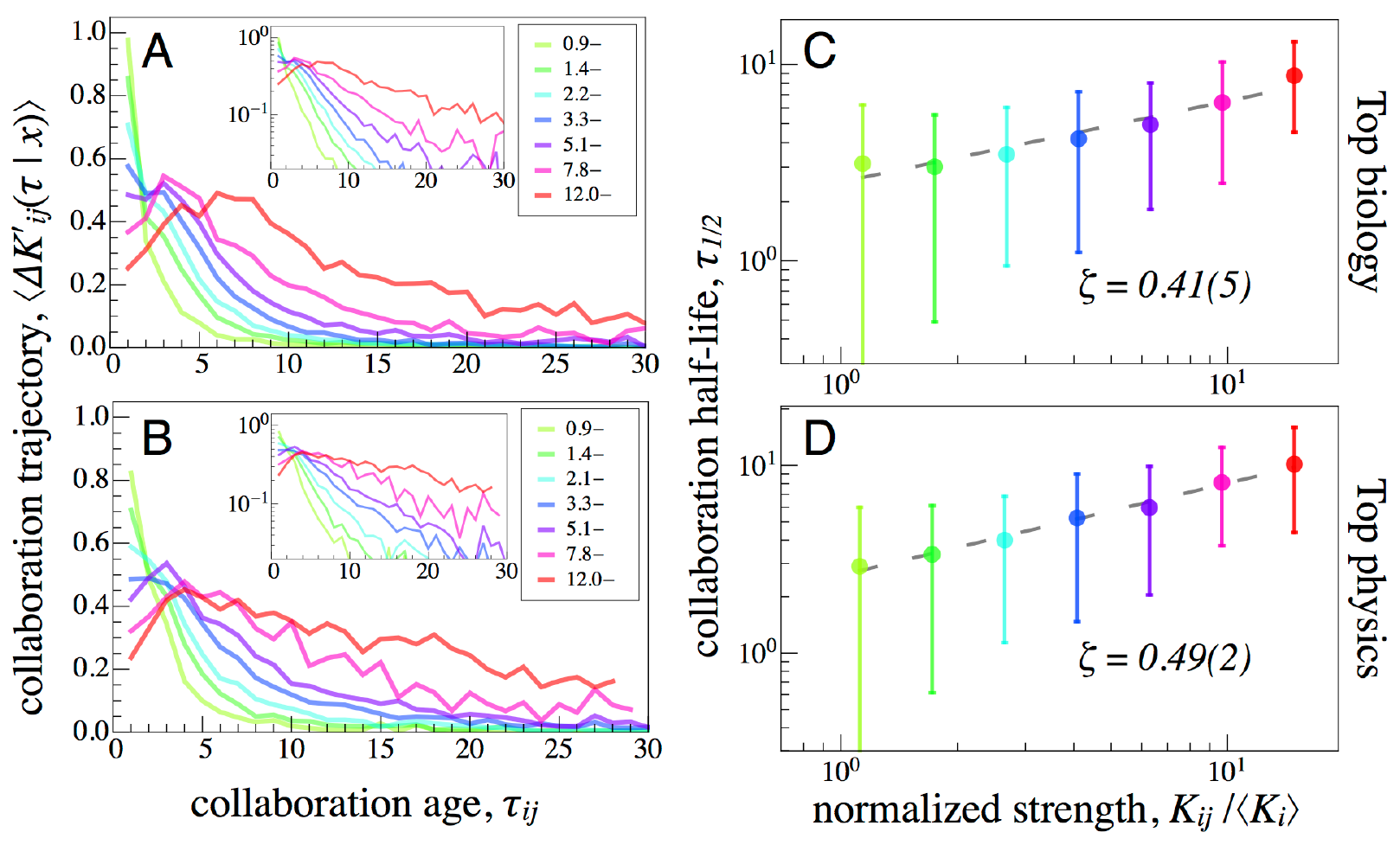}}
\caption{\label{collabtraj}  {\bf Growth and decay of collaboration ties.} (A,B) Average collaboration intensity, normalized to peak value, measured $\tau_{ij}$ years after the initiation of the collaboration tie. (Insets) On log-linear axes the  decay appears as linear, corresponding to  an exponential form.  (C,D) For each $\{x\}$ group we show the average and standard deviation (error bar) of $\tau_{1/2}$; we  use  logarithmically spaced  $\{x\}$ groups that correspond by color to the same $\{x\}$ as in panels (A,B). 
The $\zeta$ value quantifies the scaling of $\langle \tau_{1/2} \rangle$ as a function of the normalized coauthor strength $x_{ij} \equiv K_{ij}/\langle K_{i} \rangle$. The sub linear  ($\zeta<1$) values indicate that  collaborations are distributed over a timescale that grows slower than proportional to $x$; conversely, this means that longer collaborations are relatively  more productive, being characterized by increasing marginal returns ($1/\zeta>1$).
% Trajectories are truncated when there are less than 20 values to average.
{\it SI Appendix} Fig. S3 shows the analogous plot for the other physics and biology datasets; all 4 datasets exhibit similar features.
% Np THRESHOLD implicit. 
}
\end{figure*}

Establishing statistical regularities across research profiles requires the use of  a normalized duration measure, $\Delta_{ij} \equiv L_{ij}/\langle L_{i} \rangle$, which controls for author-specific collaboration patterns by measuring time in units of  $\langle L_{i} \rangle$.  
The empirical distributions are right-skewed, with approximately $63\%$ of  the data with  $L_{ij}<\langle L_{i} \rangle$ (corresponding to $\Delta_{ij} < 1$).  Nevertheless, approximately 1\% of collaborations last longer
than $4\langle L_{i} \rangle \approx$ 15 to 20 years. 
Moreover, Fig.~\ref{LongPDF}(A) shows  that the log-logistic 
probability density function (pdf)
\begin{equation}
\label{PDelta}
P(\Delta) =  \frac{(b/a)(\Delta /a)^{b-1}}{(1+(\Delta /a)^{b})^{2}} \ ,
\end{equation}
 provides a good fit to the empirical data over the entire range of $\Delta_{ij}$.
The log-logistic (Fisk) pdf is a well-known survival analysis distribution with  property Median$(\Delta)=a$.
%, and the convenient  property that the  Gini inequality coefficient $G(\Delta)=1/b$. 
By construction,  the mean value $\langle \Delta \rangle \equiv 1$, which reduces our parameter space to just $b$ as   $a=\sin(\pi/b) / (\pi/b)$.
For each dataset we calculate $b\geq 2.6$, estimating the parameter using ordinary least-squares. 
Associated with  each $P(\Delta)$ is a hazard function  representing the likelihood that a collaboration terminates for a given $\Delta_{ij}$. Since   $b>1$, the hazard function is unimodal, with a  maximum value occurring at $\Delta_{c}=a(b-1)^{1/b}$ with bounds  $\Delta_{c}>a$ for $b>2$ and $\Delta_{c}>1$ for $b>2.83...$; using the best-fit $a$ and $b$ values we estimate $\Delta_{c}\approx$ 0.94 (top biology), 1.11 (other biology), 0.77 (top physics), and 1.08 (other physics). 
%These peak values are around unity, meaning that the mean duration $\langle L_{i} \rangle$ also represents a characteristic  tipping point in the sustainability of collaboration ties since the likelihood of sustaining a collaboration reaches a minimum at $\Delta_{c}$, increasing thereafter.
Thus, $\Delta_{c}$ represents a tipping point in the sustainability of a collaboration, because the likelihood that a collaboration terminates peaks at $\Delta_{c}$ and then decreases monotonically for $\Delta_{ij} > \Delta_{c}$. This observation lends further significance to the author-specific time scale $\langle L_{i} \rangle$.
The  log-logistic  pdf is also characterized by  asymptotic power-law behavior $P(\Delta) \sim \Delta ^{-(b+1)}$ for large $\Delta_{ij}$. 

In order to determine how the  $\Delta_{ij}$ values are distributed across the career, we calculated the mean duration $\langle \Delta \vert t \rangle$ using a 5-year (sliding window) moving average centered around career age $t$. If the $\Delta_{ij}$ values were distributed independent of $t$, then  $\langle \Delta \vert t \rangle \approx 1$. 
Instead, Figure~\ref{LongPDF}(B) shows a negative trend for each dataset. 
 Interestingly, the  $\langle \Delta \vert t \rangle$ values are consistently larger for the top scientists, indicating that the relatively  
short $L_{ij}$ are more concentrated at larger $t$. This pattern of  increasing access to short-term collaboration opportunities  points to an additional positive feedback mechanism contributing to cumulative advantage \cite{petersen_quantitative_2011,petersen_inequality_2014}.\\

%It is also important to know the distribution of $L_{ij}$, which provides important insights into the turnover rate of collaborations in science.
%To this end, we defined a normalized duration, $\Delta_{ij}$, that is better for aggregating data across  researcher profiles characterized by varying $\langle L_{i} \rangle$. We find that the entire  $P(\Delta)$  distribution can be modeled by the 2-parameter  log-logistic function.  Furthermore, by knowing the functional form of $P(\Delta)$, it is  possible to calculate the likelihood (hazard rate) that a collaboration terminates as a function of $\Delta_{ij}$. For the log-logistic  parameters we estimated,  the hazard rate is characterized by a single peak around $\Delta_{c}\approx 1$ (corresponding to $L_{ij} \approx \langle L_{i} \rangle$). 
\noindent{\bf Quantifying the collaboration life cycle.}
The $P(\Delta)$ distribution points to the  variability of  time scales in the scientific collaboration  network -- while a small number of collaborations last a lifetime, the remainder decay quite quickly in a collaboration environment characterized by a remarkably high churn rate. 
 %growth patterns of  $K_{ij}(t)$,  the cumulative number of publications between $i$ and $j$ up to year $t$. 
 Since  it  is possible that  a relatively long $L_{ij}$ corresponds to just the minimum 2 publications, it is also important to analyze the collaboration rate. To this end, we quantify the patterns of growth and decay in tie strength using the more than 166,000  dyadic $(ij)$ collaboration records: $K_{ij}(t)$ is  the cumulative number of coauthored publications  between  $i$ and $j$ up to year $t$, and  $\Delta K_{ij}(t)=K_{ij}(t)-K_{ij}(t-1)$ is the annual  publication rate.

In order to define a collaboration trajectory that is better suited for
averaging, we normalize each individual $\Delta K_{ij}(\tau)$ by its peak value, 
  \begin{equation}
  \Delta K_{ij}'(\tau)\equiv \Delta K_{ij}(\tau)/{\rm  Max}[\Delta K_{ij}(\tau)] \ .
   \end{equation} 
Here    $\tau \equiv \tau_{ij} = t-t_{ij}^{0}+1$ is the number of years since the initiation of a given collaboration.
  %This quantity is better suited for understanding the overall dynamics of $\Delta K_{ij}(t)$, independent of its amplitude. 
This normalization procedure is useful for comparing and averaging time series' that are characterized by just a single peak.  

Expecting that the collaboration trajectories depend on the  tie strength, we  grouped the individual  $ \Delta K_{ij}'(\tau)$  according to the normalized coauthor strength, $x_{ij} \equiv K_{ij}/\langle K_{i} \rangle$. The normalization factor $\langle K_{i} \rangle = S_{i}^{-1} \sum_{j=1}^{S_{i}} K_{ij}$ is calculated across the $S_{i}$ distinct collaborators (the collaboration radius of $i$),  and represents an intrinsic collaboration scale which grows in proportion to both an author's typical collaboration size and his/her publication rate.  We then aggregated the $N_{\{x\}}$ trajectories in each  $\{x\}$ group and  calculated the average trajectory
  \begin{equation}
 \langle \Delta K_{ij}'(\tau | x) \rangle \equiv N_{\{x\}}^{-1} \sum_{\{x\}} \Delta K_{ij}'(\tau| x) \ .
 \end{equation} 

%Thus, we measured how $\Delta K_{ij}(t)$, the number of publications coauthored by $i$ and $j$ in career year $t$,  depends on the tie strength, $K_{ij}$.  
%(see the {\it SI Text}  for the details of this additional analysis).
Indeed, Fig.~\ref{collabtraj} shows that the collaboration `life cycle'  $\Delta K_{ij}(\tau | x)$ depends strongly on the relative tie strength $x_{ij} \equiv K_{ij}/ \langle K_{i} \rangle$.  The trajectories with  $x_{ij} >12.0$ decay over a relatively long timescale, maintaining a value approximately $0.2\ {\rm  Max}[\Delta K_{ij}(\tau)]$ even 20 years after initiation, reminiscent of  a `research life partner'. The trajectories with $x_{ij}\in [0.9, 1.4]$ represent  common  collaborations that decay exponentially over the characteristic  time-scale $\langle L_{i} \rangle$. A mathematical side note, useful as a modeling benchmark, is the  linear decay when plotted on log-linear axes, suggesting a functional form that is  exponential  for large $\tau$, $\langle \Delta K_{ij}'(\tau | x) \rangle \sim \exp[-\tau/\overline{\tau}]$.

We further emphasize the ramifications of the life-cycle variation by quantifying  the relation between  $x_{ij}$ and the collaboration's  half-life $\tau_{1/2}$, defined as  the number of years to reach half of the total collaborative output according to the relation $K_{ij}(t=\tau_{1/2})=K_{ij}/2$. We observe  a   scaling relation  $\langle \tau_{1/2} \rangle \sim x^{\zeta}$ with $\zeta$ values ranging from 0.4 to 0.5. 
Sublinear values  ($\zeta<1$) indicate  that a collaboration with twice the strength is likely to have a corresponding $\tau_{1/2}$ that is  less-than doubled. 
This feature captures the burstiness of collaborative activities, which  likely arises from the heterogenous  overlapping of  multiple timescales, e.g. the variable  contract lengths in science ranging from single-year contracts to lifetime tenure, the overlapping of multiple age cohorts, and the projects and grants  themselves which are typically characterized by  relatively short terms. 
 Nevertheless, $dx/d\tau_{1/2} \sim \tau_{1/2}^{(1-\zeta)/\zeta}$ is increasing function for $\zeta<1$, indicating an increasing marginal returns with  increasing  $\tau_{1/2}$, further signaling  the productivity benefits of long-term collaborations characterized by formalized roles, mutual trust, experience,  and group  learning that together facilitate efficient interactions.\\
%{\color{blue} Technical and social trust -- Centralized Ecosystem model only works if there are clear and fair rules on the handling of the input and output; Contracts are important for the management of the outcome, and are good for establishing trust -- point out that in science careers the contracts are more social, relying on ethical behavior.}

\begin{figure}
\centering{\includegraphics[width=0.45\textwidth]{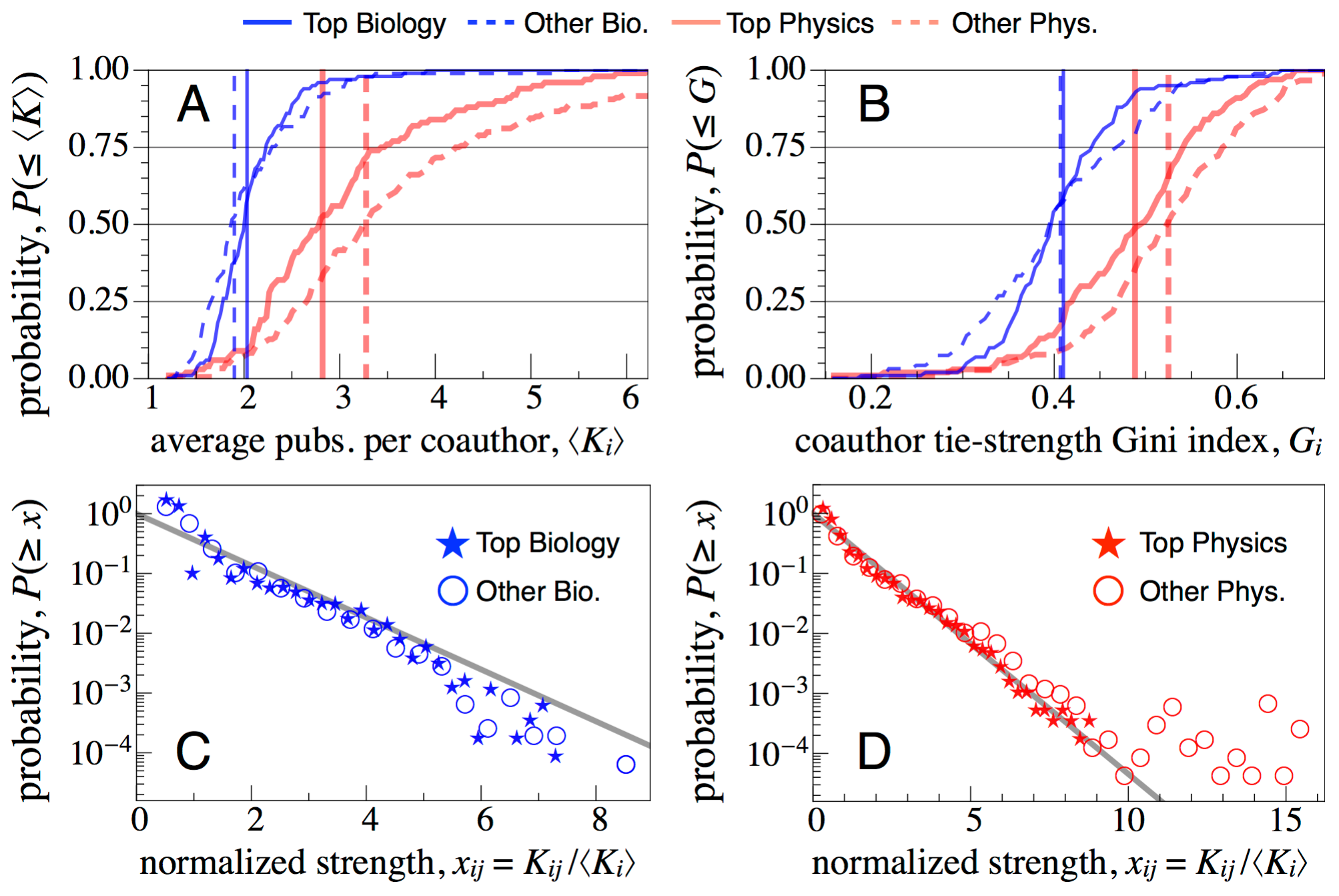}}
\caption{\label{TieDist} {\bf Characteristic measures of collaboration tie strength.} (A) Cumulative distribution of the mean collaboration strength, $\langle K_{i} \rangle$. The Kolmogorov-Smirnov (K-S) test indicates that the  $P(\langle K_{i} \rangle)$ are similar  for biology ($p=0.031$) and significantly different for physics ($p=0.004$). Vertical lines indicate median value. (B) Cumulative distribution of $G_{i}$. The pairwise K-S test indicates that the $P(G_{i})$ are similar for biology ($p = 0.14$) but not for physics ($p = 0.02$). Vertical lines indicate the mean value, with physics  indicating significantly higher $G_{i}$ than for biology. (C,D) For each dataset, the cumulative distribution of  normalized collaboration strength $x_{ij}$ shows excellent agreement with the exponential distribution $E(x) = \exp[-x]$ (gray line) over the bulk of the distribution, with the deviations in the tail regime representing less than 0.1\% of the data.  
}
\end{figure}

\noindent{\bf Quantifying the tie-strength distribution.} 
Here we focus on the cross-sectional distribution of tie strengths within the ego network. We use  the final tie strength value $K_{ij}$ to   
 distinguish the strong ties ($K_{ij} \geq \langle K_{i} \rangle$) from the weak ties ($K_{ij} < \langle K_{i} \rangle$). 
Figure \ref{TieDist}(A) shows the cumulative distribution $P(\leq \langle K_{i} \rangle)$ of the mean tie strength $ \langle K_{i} \rangle$,  which can vary over a wide range   depending on a researcher's involvement in large team science activities. We also quantify the concentration of tie strength  using the  Gini index $G_{i}$ calculated from each researcher's $K_{ij}$ values; the distribution $P(\leq G_{i})$ is shown in Fig. \ref{TieDist}(B). Together, these two measures capture the variability in collaboration strengths across and within discipline, with physics exhibiting larger $\langle K_{i} \rangle$ and $G_{i}$ values. 
%The distributions also indicate that the biology datasets are well matched with respect to these two quantities, whereas the physics datasets are less well matched, with smaller values for the top cited researchers. Hence, we will control for this variability  in our ultimate regression model. %While there are  many factors that affect the variation in $\langle K_{i} \rangle$, we leave this for an independent study. 
%Despite the variation in $\langle K_{i} \rangle$, we find that the distribution $P(K_{ij})$   exhibits common features, ranging rather smoothly from the collaborations with the smallest value $K_{ij}=1$ (the most frequent)  to the large $K_{ij}$ which are limited by the upper bound $K_{ij} \leq N_{i}$.\\

Another important author-specific variable is the publication overlap between each researcher and his/her top collaborator. 
This measure is defined as the fraction of a researcher's $N_{i}$ publications including his/her top collaborator, $f_{K,i} = Max_{j}[K_{ij}]/N_{i}$.
We observe surprisingly large variation in  $f_{K,i}$, with  mean and standard deviation in the range of $ 0.16 \pm 0.14$ for the top scientists and $0.36 \pm 0.23$ for the other scientists. Across all profiles, the min and max  $f_{K,i}$ values  are  $0.03$ and $0.99$, respectively, representing nearly the maximum possible  variation in observed publication overlap. 
An example of this limiting scenario is shown in  Fig. S2, highlighting the ``dynamic duo''  of J. L. Goldstein and M. S. Brown, winners of the 1985 Nobel Prize in Physiology or Medicine; Goldstein and Brown  published more than 450 publications each, with roughly  $100 \times f_{K,i} \approx 95\%$ coauthored together. Remarkably, we find that overlaps larger than 50\% are  not uncommon, observing   $100P(f_{K}\geq 0.5) \approx 9\%$  (biology) and $100P(f_{K}\geq 0.5) \approx 20\%$ (physics) of $i$ having more than half of their publications with their strongest collaborator. 

 However, within a researcher profile, it is likely that more than just the top collaborator was central to his/her career. Indeed, key to our investigation is the  identification of the extremely strong collaborators --  super ties -- that are distinguished within the subset of strong ties.
%Recently the  ``h-index'' concept has been extended to the coauthor distribution as a method to distinguish the core collaborators \cite{ausloos_scientometrics_2013}. IF THERE IS ROOM
% The remainder of this study involves identifying and analyzing the relative impact of the super ties within each collaboration profile. Here
 Hence, using the empirical information contained within each researcher's  tie-strength  distribution, $P(K_{ij})$, we develop an objective super-tie criteria that is author-specific.
%across our sample by defining  a normalized collaboration strength $x_{ij} \equiv K_{ij}/\langle K_{i} \rangle$ which is better suited for aggregating across scientists.
First, in order to gain a better understanding of the statistical distribution of $K_{ij}$, we  aggregated the tie-strength data across all research profiles, using the  normalized collaboration strength  $x_{ij}$.
Figures \ref{TieDist}(C,D) show the cumulative distribution $P(\geq x)$ for each discipline. Each $P(\geq x)$ is in good agreement with the exponential distribution $\exp[-x]$ (with mean value $\langle x \rangle = 1$ by construction), with the exception in the  tail, $P(\geq x) \lesssim 10^{-3}$,  which is home to extreme collaborator  outliers. Thus, by a second means in addition to the result for $L_{ij}$, we find that roughly 2/3 of the ties we analyzed are weak (i.e. the fraction of observations with $x_{ij}<1$ is given by $1-1/e \approx 0.63$). 

Based upon this empirical evidence, we use  the discrete exponential distribution as our  baseline model, $P(K_{ij}) \propto \exp(-\kappa_{i} K_{ij})$. We then use extreme statistics arguments to precisely define the author-specific super-tie threshold $K^{c}_{i}$. The extreme statistic criteria posits   that out of the $S_{i}$ empirical observations there should be just a single observation with $K_{ij} > K^{c}_{i}$.   The threshold $K^{c}_{i}$ is operationalized by integrating  the tail of  $P(K_{ij})$ according to the equation $1/S_{i} = \sum_{K_{ij}>K^{c}_{i}}^{\infty}P(K_{ij})=\exp(-\kappa_{i} K^{c}_{i})$, with the analytic relation $\langle K_{i} \rangle = \sum_{K_{ij}=1}^{\infty} K_{ij}P(K_{ij}) =e^{\kappa_{i}}/(e^{\kappa_{i}}-1) \approx 1+1/\kappa_{i}$ for small $\kappa_{i}$. 
 In the relatively large $S_{i}$ limit, $K^{c}_{i}$ is given by the simple relation
\begin{equation}
K^{c}_{i} = (\langle K_{i} \rangle-1) \ln S_{i} \ .
\label{Kc}
\end{equation}
The advantage of this approach is that $K^{c}_{i}$ is nonparametric, depending only on the observables   $\langle K_{i} \rangle$ and $S_{i}$.
Thus, the super-tie threshold is proportional to $\langle K_{i} \rangle-1$ (the $-1$  arises because the minimum $K_{ij}$ value is 1),  with a logarithmically factor  $ \ln S_{i}$ reflecting the sample size dependence. 
This extreme value criteria is  generic, and can be derived for any data following  a  baseline distribution; for a succinct explanation of this analytic method see page 17 of ref. \cite{krapivsky_kinetic_2010}.

In what follows, we label each coauthor $j$ with $K_{ij}>K^{c}_{i}$ a super  tie, with indicator variable $R_{j}\equiv1$. The rest of the ties with $K_{ij}\leq K^{c}_{i}$ have an indicator variable  $R_{j}\equiv0$.  This method has limitations, specifically in the case that the collaboration profile  does not follow an exponential $P(K_{ij})$. For example, consider the extreme case where every $K_{ij}=1$, meaning that  $K^{c}_{i} =0$ (independent of $S_{i}$), resulting in all coauthors being super ties ($R_{j}=1$ for all $j$). This scenario is rare and unlikely to occur for  researchers with relatively large $N_{i}$ and $S_{i}$, as in our researcher sample.\\

\begin{figure}
\centering{\includegraphics[width=0.45\textwidth]{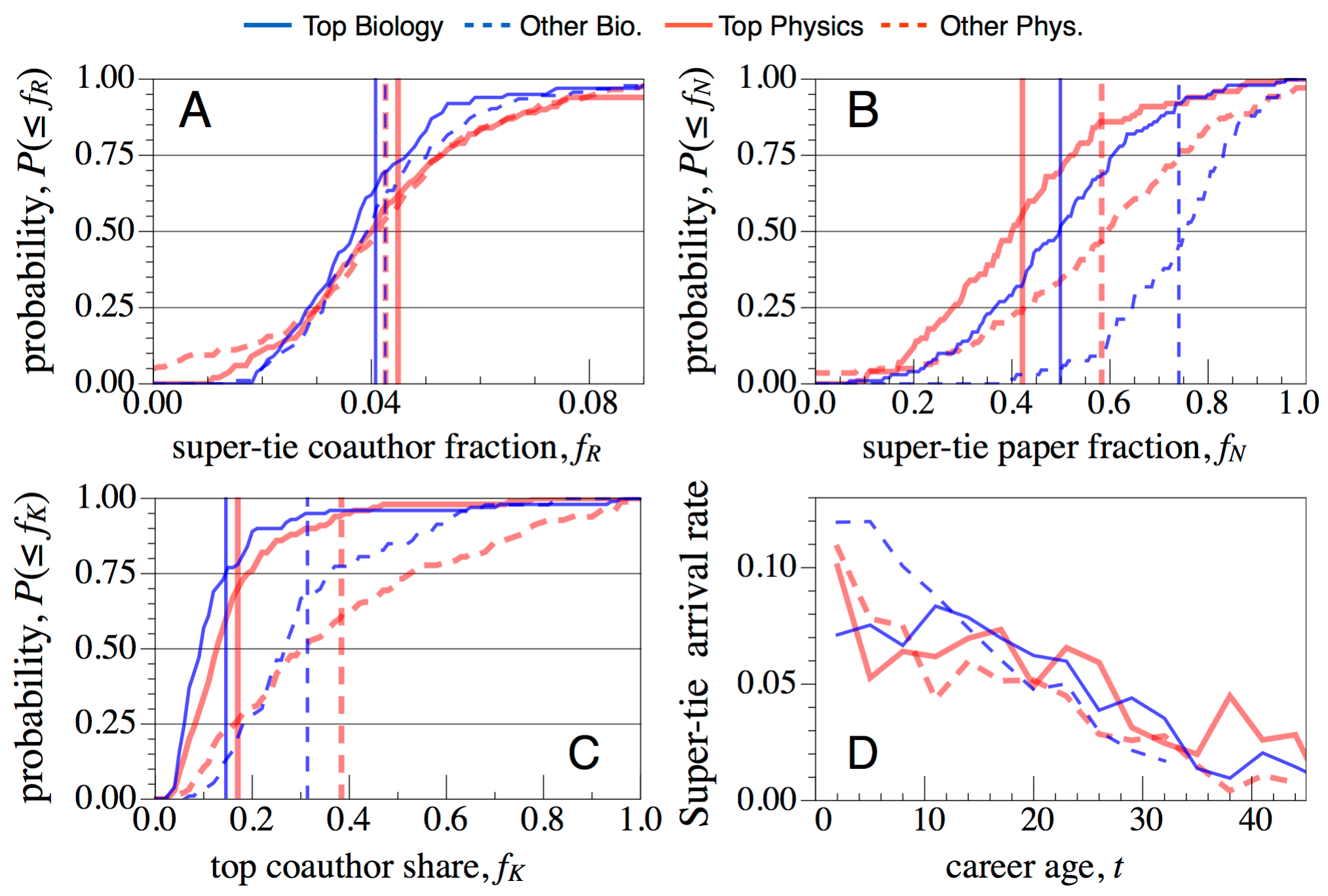}}
\caption{\label{SuperTieDist} {\bf The frequency  of super ties.}  Vertical lines indicate the distribution mean.  (A) Cumulative distribution of the fraction $f_{R,i}$ of the $S_{i}$ coauthors that are super ties. All pairwise comparisons of the distributions have K-S $p$-value greater than 0.21 indicating a common underlying  distribution $P(f_{R})$. (B) Cumulative distribution of the fraction $f_{N,i}$ of publications that include at least one super-tie coauthor. 
The top scientist distributions show  mean values that are significantly smaller than their counterparts.  (C) Cumulative distribution of the fraction $f_{K,i}$ of publications coauthored with his/her top collaborator. The mean  and standard deviation for  biology (top) is   $0.15\pm0.16$, for biology (other) is $0.31\pm 0.16$, for physics (top) is $0.17\pm0.13$, and for physics (other) is $0.38\pm 0.26$. (D) The mean rate of super-ties per new collaboration, $\langle \lambda_{R}(t)\rangle$, averaged over all the profiles in each dataset using observations aggregated over consecutive 3-year  periods.  
}
\end{figure}

\noindent{\bf Quantifying the prevalence and impact of super ties.} 
How common are super ties? For each profile we denote the number  of coauthors that are super ties by  $S_{R,i}$  (with complement $S_{!R,i}=S_{i}-S_{R,i}$).  {\it SI Text} Fig. S4 shows that the distribution of $S_{R,i}$  is rather broad, with mean  and standard deviation  $S_{R,i}$ values: $18\pm13$ (top bio.), $16\pm 13$ (other bio.),  $7.3\pm4.8$ (top phys.),  $6.8\pm 5.1$ (other phys.).
% Comparison top-other  $P(S_{R})$ within discipline yields K-S test statistic p-value 0.123966 (bio) and 0.335993 (phys), meaning the datasets are well-matched.
 The super-tie coauthor fraction, $f_{R,i}=S_{R,i}/S_{i}$, measures the super-tie frequency on a  per-collaborator basis, with  mean value  $ \langle f_{R} \rangle \approx 0.04$ (i.e. typically 1 super tie for every 25 coauthors). Furthermore, Fig. \ref{SuperTieDist}(A) shows that the distribution $P(\leq f_{R})$  is common across the four datasets. 
We tested the universality of the probability distribution $P(f_{R})$ between the top and other researcher datasets using  the  Kolmogorov-Smirnov (K-S)  statistic, which tests the null hypothesis that the data come from the same underlying pdf.
The smallest pairwise K-S test $p$-value between any two $P(f_{R})$ is  $p=0.21$, indicating that  we fail to reject the null hypothesis that  the distributions are equal, highlighting  that the four datasets are remarkably well-matched with respect to the distribution of $f_{R,i}$.

On a per paper basis, Fig. \ref{SuperTieDist}(B) shows that the fraction  of a researcher's  portfolio coauthored with at least one super tie, $f_{N,i}$,  can vary over the entire range of possibilities, with mean  and standard deviation  $0.50\pm0.18$ (top bio.), $0.74\pm 0.13$ (other bio.),  $0.42\pm0.19$ (top phys.),  $0.58\pm 0.23$ (other phys.).  
Furthermore, we found that  41\% of  the top scientists have $f_{N,i}\geq 0.5$.
%This feature is further reiterated by considering the publication overlap with the researcher's strongest tie, $f_{K,i}$, shown in Fig. \ref{SuperTieDist}(C). 
Interestingly, the distributions of  $f_{K,i}$ and $f_{N,i}$  indicate that top scientists have lower levels of  super-tie dependency than their counterparts.

We also analyzed the arrival rate  of  super-ties. For each profile we tracked the number of super ties initiated in year $t$, and normalized this number by the total number of new collaborations initiated in the same year. This ratio, $\lambda_{R,i}(t)$,   estimates the likelihood that a new collaboration eventually becomes a super tie as a function of  career age $t$.  For example, using the set of collaborations initiated in each scientist's first year, we estimate the  likelihood that a first-year collaborator (mentor)   becomes a super tie at  $\lambda_{R}(t=1) = 8\%$ (top bio.), $16\%$ (other bio.), $14\%$ (top phys.), and $15\%$ (other phys.). 
Figure \ref{SuperTieDist}(D) shows the mean arrival rate, $\langle \lambda_{R}(t) \rangle$, calculated by averaging over all profiles in each dataset. The super tie arrival rate  declines across the career, reaching a 5\% likelihood per new collaborator at $t=20$ and 2.5\% likelihood by $t=30$. The decay is not as fast for the top-cited scientists, possibly reflecting their preferential access to outstanding collaborators. However, the estimate for large $t$  is  biased toward smaller values because  collaborations initiated late in the career  may not have had sufficient time to grow. 

In the next two subsections, we investigate the role of super ties at the micro level by analyzing productivity at the annual time resolution and the citation impact  of individual publications.
In the {\it SI Text} we provide additional evidence for the advantage of super  ties by developing  descriptive methods that  measures the net productivity and citations of the super ties relative to all other ties. \\

\noindent{\bf The Apostle effect I: Quantifying  the impact of super ties on annual productivity.}  
We analyzed each research profile over the career years $t_{i} \in [6,Min(29,T_{i})]$,  separating the data into non-overlapping $\Delta t$-year periods, and neglecting the first 5 years to allow the $L_{ij}(t)$ and $K_{ij}(t)$ sufficient time to grow. 
We then modeled the dependent variable, $n_{i,t}/\langle n_{i} \rangle$, which is the  productivity aggregated over  $\Delta t$-year periods, normalized by the baseline average calculated over the period of analysis.  Recent analysis of assistant and tenured professors has shown  that  the annual publication rate  is governed by slow but substantial growth across the career, with fluctuations that are largely related to collaboration size \cite{petersen_persistence_2012}.

To better understand the factors contributing to productivity growth, we include controls for career age $t$ along with four additional  variables measuring the composition of  collaborators from each $\Delta t$-year period. First, we calculated the average number of authors per publication, $\overline{a}_{i,t}$,  a proxy for labor input, coordination costs, and the research technology level.  Second, we calculated the mean  duration, $\overline{L}_{i,t}$,   by averaging the   $L_{ij}(t-\Delta t)$ values (from the previous period) across only the $j$ who  are active in $t$ -- i.e. those coauthors with $\Delta K_{ij}(t) > 0$. 
 In this way, we account for the possibility that $j$ was not active in the previous period $(t-\Delta t)$, in which case $L_{ij}(t-\Delta t)$ is even smaller than  $L_{ij}(t)-\Delta t$. Thus,  $\overline{L}_{i,t}$ measures the  prior experience between $i$ and his/her collaborators. Third, for the same set of coauthors as for $\overline{L}_{i,t}$, we calculated  the Gini index of the collaboration strength, $G^{K}_{i,t}$,  using  the tie strength values up to the previous period, $K_{ij}(t-\Delta t)$. Thus, $G^{K}_{i,t}$ provides a standardized measure of the dispersion in coauthor activity, with values ranging from 0 (all coauthors published equally in the past with $i$) to 1 (extreme inequality in prior  publication with $i$).
Thus, while  $\overline{L}_{i,t}$ measures the lifetime  of the group's prior collaborations,  $G^{K}_{i,t}$ measures the  concentration of their prior experience.
  And finally, for each period  $t$, we calculated the contribution  of  super tie collaborators  normalized by the  contribution of all other collaborators,
\begin{equation}
\rho_{i,t}\equiv \frac{\sum_{j\vert R=1} \Delta K_{ij}(t)}{\sum_{j\vert R=0}\Delta K_{ij}(t)} \ ,
\label{rhoeq}
\end{equation}
accounting for the possibility that the relative contribution of super ties may affect productivity.
While the total coauthor contribution $\sum_{j}\Delta K_{ij}(t)$ is highly correlated with $n_{i,t}$, the correlation coefficient between  $\rho_{i,t}$ and $n_{i,t}$ is only 0.07.
We only include researchers in this analysis if  there are $\geq4$ data points  for which the denominator of Eq.~[\ref{rhoeq}] is nonzero. 

We implemented a fixed effects regression of the model
\begin{eqnarray}
\frac{n_{i,t}}{\langle n_{i} \rangle} &=& \beta_{i,0} + \beta_{\overline{a}} \ln \overline{a}_{i,t} + \beta_{\overline{L}} \overline{L}_{i,t}  + \nonumber \\
& &  \beta_{G} G^{K}_{i,t}+ \beta_{\rho} \rho_{i,t} +\beta_{t} t_{i,t}+ \epsilon_{i,t} \ ,
\label{Reg1}
\end{eqnarray}
%in STATA11 using  ``xtreg , vce(robust) fe'' to calculate
which accounts for author-specific time-invariant features ($\beta_{i,0}$), using  robust standard errors to account for  autocorrelation within each $i$. 
 Because the predictors are calculated from the same ego profile, covariance is expected; for example, the highest correlation coefficient between any two independent variables is $0.32$ between $\ln \overline{a}_{i,t}$ and $G^{K}_{i,t}$, because the  variance in $K_{ij}$ increases proportional to the sample size (i.e. $\overline{a}_{i,t})$.
Table 1 shows the results of our model estimates for $\Delta t=1$ year and Table S1 shows the results for $\Delta t =3$ years. We also ran the regression for all the datasets together,``All'', and provide standardized  coefficients that better facilitate a comparison of the coefficient magnitudes.
 
We observed a positive coefficient $\beta_{\rho}=0.11 \pm 0.01$   ($p\leq 0.003$ for all datasets), meaning that larger contributions by super ties is associated with above-average productivity.  By way of example, consider a scenario where the super ties contribute a  third of the total coauthor input, corresponding to $\rho_{i,t} = 0.5$,  the average $\rho_{i,t}$ value we observed. Consider a second scenario with $\rho_{i,t} = 1$, corresponding to equal input by the super ties and their counterparts ($\rho_{i,t} \geq1$ for 14\% of the observations). 
If all other parameters contribute  a baseline productivity value 1, then the additional  contribution from $\beta_{\rho}$ corresponds to a $100 \times 0.5\beta_{\rho}/(1+ 0.5\beta_{\rho}) = 5.2$\%  productivity increase. 
This value is consistent with the $5\%$ productivity spillover observed in a study of star scientists \cite{azoulay_superstar_2010}.

 %  ALL DOES NOT INCLUDE DATASET/DISCIPLINE DUMMIES BECUASE THEY ARE OMMITED AS COLLINEAR VARIABLE BY STATA
 We also found that periods corresponding to higher levels of prior experience are associated with below-average productivity ($\beta_{\overline{L}}<0$, $p\leq 0.008$ for all datasets except for top biology). Despite the costs associated with tie-formation, this result demonstrates that productivity can benefit from collaborator turnover. Nevertheless, above-average productivity is associated with higher inequality in the concentration of prior experience ($\beta_{G}>0$, $p< 0.001$ level for all datasets). Together, these results point to the benefits of strategically pairing  new collaborators with incumbent ones in order to  
promote the  atypical combination of knowledge backgrounds and to achieve higher scientific impact  \cite{Uzzi_Atypical_2013}.  In Table 1 we also report standardized coefficients that facilitate  a comparison of the relative strengths of the model variables, revealing that $\beta_{G}$ is twice as strong as $\beta_{\rho}$ and $\beta_{\overline{L}}$. 
Interestingly, $\beta_{\rho}$ and $\beta_{\overline{L}}$ have  opposite signs, yet  are balanced in magnitude, suggesting a compensation strategy for group managers.
  
The age coefficient $ \beta_{t}$ is also positive ($p< 0.001$ level for all datasets), consistent with patterns of steady productivity growth observed for successful research careers  \cite{petersen_reputation_2014,petersen_persistence_2012,petersen_inequality_2014}.  Possible  explanatory variables to consider in extended analyses are the standard deviation in $K_{ij}$, a contact frequency  ($K_{ij}/L_{ij}$) measure of tie strength intensity per  Granovetter's original operationalization \cite{granovetter_strength_1973}, and absolute calendar year $y$, variables which we omit here to keep the model streamlined.\\

%TABLE HERE
\begin{table*}
\caption{Parameter estimates for  the productivity model  in Eq. (6) using $\Delta t = 1$ year long periods, and the citation model in Eq. (8) using only the publications with $y_{p}\leq 2003$. Each fixed effects model was calculated using robust standard errors,   implemented by the Huber/White/sandwich method. 
%See Table \ref{table:regprod2} for robustness check with  $\Delta t=3$. 
%These results demonstrate the robustness of the model with respect to $\Delta t$.
%For the productivity model, only profiles with 4 or more data values were included in the regression.
%For the citation model, only publications with $y_{p}\leq 2002$  were analyzed so that the dependent variable $z_{i,p}$ has  sufficient time to become a  robust measure of citation impact.
Values significant at the $p\leq0.04$ level are indicated in boldface. ``Std. coeff.'' represents the estimates of the standardized (beta) coefficients.%, except for $t$, $R_{p}$, and $t_{p}$.
``All'' corresponds to the combination of all datasets.
% USE GINI INDEX (instead of kiave and kids) SINCE IT IS LESS CORRELATED WITH \rho then the other two: correlate liave_t1 kiave_t1 kisd_t1 ginik_t1 krbykr_t
}
%\small
%\resizebox{1.65\columnwidth}{!}{
\begin{tabular}{@{\vrule height 10.5pt depth4pt  width0pt}lc|c|c|c|c|c|c|c|}\\
&\multicolumn1l{}&\multicolumn7c{Apostle effect I: productivity model  ($n_{i,t}$) %with $\Delta t = 1$
}\\
\hline
\noalign{
\vskip-0pt}
\vrule depth 6pt width 0pt \textbf{\em Dataset}  &   $A$  & $\ln \overline{a}_{t}$ &  $\overline{L}_{t}$ &$G^{K}_{t}$   &   $\rho_{t}$ & $t$ & $N_{obs.}$ & Adj. $R^{2}$ \\
\hline  
\hline  
All & 466 & $0.002  \pm  0.029 $ & ${\bf -0.054 \pm 0.008} $ & ${\bf 1.788 \pm 0.134 }$ & ${\bf 0.110 \pm 0.013 }$ & ${\bf 0.029 \pm 0.002 }$ & $ 8483 $ & 0.19 \\
(Std. coeff.)&  & $0.002  \pm  0.033 $ & ${\bf -0.140 \pm 0.021 }$ & ${\bf 0.320 \pm 0.024 }$ & ${\bf 0.140 \pm 0.016 }$ & ${\bf 0.049 \pm 0.004 }$ &  &  \\ 
$p$-value  &  & $ 0.943 $ & $ 0.000 $ & $ 0.000 $ & $ 0.000 $ & $ 0.000 $ &  &  \\
Biology (top) & 99 & ${\bf -0.123  \pm  0.056}$ & $ -0.011 \pm 0.018 $ & ${\bf 2.816 \pm 0.270 }$ & ${\bf 0.111 \pm 0.026 }$ & ${\bf 0.031 \pm 0.003 }$ & $ 2202 $ & 0.24 \\ $p$-value  &  & $ 0.031 $ & $ 0.519 $ & $ 0.000 $ & $ 0.000 $ & $ 0.000 $ &  &  \\
Biology (other) & 95 & $-0.061  \pm  0.056 $ & ${\bf  -0.067 \pm 0.025 }$ & ${\bf  1.654 \pm 0.287 }$ & ${\bf  0.071 \pm 0.023 }$ & ${\bf  0.053 \pm 0.006 }$ & $ 1467 $ & 0.29 \\
 $p$-value  &  & $0.275 $ & $ 0.008 $ & $ 0.000 $ & $ 0.003 $ & $ 0.000 $ &  &  \\
 Physics (top) & 100 & ${\bf  -0.146  \pm  0.057 }$ & ${\bf   -0.047 \pm 0.015 }$ & ${\bf   2.053 \pm 0.287 }$ & ${\bf   0.153 \pm 0.025 }$ & ${\bf   0.022 \pm 0.004 }$ & $ 2056 $ & 0.15 \\ $p$-value  &  & $ 0.012 $ & $ 0.002 $ & $ 0.000 $ & $ 0.000 $ & $ 0.000 $ &  &  \\
 Physics (other) & 172 & $0.089  \pm  0.050 $ & ${\bf -0.065 \pm 0.013 }$ & ${\bf 1.495 \pm 0.213 }$ & ${\bf 0.101 \pm 0.021 }$ & ${\bf 0.026 \pm 0.005 }$ & $ 2758 $ & 0.15 \\
  $p$-value  &  & $ 0.079 $ & $ 0.000 $ & $ 0.000 $ & $ 0.000 $ & $ 0.000 $ &  &  \\
 \hline
 \noalign{\vskip5pt}
 &\multicolumn1l{}&\multicolumn7c{Apostle effect II: citation  model ($z_{i,p}$)}\\
\hline
\noalign{
\vskip-0pt}
\vrule depth 6pt width 0pt \textbf{\em Dataset}  &   $A$  & $\ln a_{p}$ &  $R_{p}$ & $t_{p}$   &   $\ln N_{i}(t_{p})$ & $\ln S_{i}(t_{p})$ & $N_{obs.}$ & Adj. $R^{2}$ \\
\hline  
\hline  
All & 377 & ${\bf 0.263  \pm  0.024}$ & ${\bf 0.202 \pm 0.023}$ & ${\bf -0.061 \pm 0.004}$ & $ 0.062 \pm 0.066 $ & $ 0.065 \pm 0.072 $ & $ 68589 $ & 0.27 \\ 
(Std. coeff.) & & ${\bf 0.135  \pm  0.012} $ & ${\bf 0.129 \pm 0.015}$ & ${\bf -0.039 \pm 0.003}$ & $0.044 \pm 0.046$ & $0.050 \pm 0.055$ &  &  \\ 
$p$-value  &  & $ 0.000 $ & $ 0.000 $ & $ 0.000 $ & $ 0.347 $ & $ 0.367 $ &  &  \\
Biology (top) & 100 & ${\bf 0.263  \pm  0.039}$ & ${\bf 0.213 \pm 0.033}$ & ${\bf -0.029 \pm 0.007}$ & $ -0.138 \pm 0.102 $ & $ 0.062 \pm 0.112 $ & $ 22135 $ & 0.12 \\
 $p$-value  &  & $ 0.000 $ & $ 0.000 $ & $ 0.000 $ & $ 0.177 $ & $ 0.578 $ &  &  \\
Biology (other) & 55 & ${\bf 0.579  \pm  0.053}$ & ${\bf 0.152 \pm 0.066}$ & ${\bf -0.031 \pm 0.015}$ & $-0.179 \pm 0.095$ & ${\bf 0.211 \pm 0.094}$ & $ 4801 $ & 0.20 \\
 $p$-value  &  & $ 0.000 $ & $ 0.026 $ & $ 0.040 $ & $ 0.065 $ & $ 0.029 $ &  &  \\
Physics (top) & 100 & ${\bf 0.139  \pm  0.043}$ & ${\bf 0.230 \pm 0.044}$ & ${\bf -0.070 \pm 0.007}$ & ${\bf 0.277 \pm 0.118}$ & $-0.119 \pm 0.135$ & $ 22673 $ & 0.19 \\
 $p$-value  &  & $ 0.002 $ & $ 0.000 $ & $ 0.000 $ & $ 0.021 $ & $ 0.380 $ &  &  \\
 Physics (other) & 122 & ${\bf 0.272  \pm  0.042}$ & ${\bf 0.235 \pm 0.049}$ & ${\bf -0.060 \pm 0.008}$ & $ 0.082 \pm 0.095 $ & $ 0.017 \pm 0.104 $ & $ 18980 $ & 0.19 \\
  $p$-value  &  & $ 0.000 $ & $ 0.000 $ & $ 0.000 $ & $ 0.389 $ & $ 0.870 $ &  &  \\
 \hline
\end{tabular}
%} %end  \resizebox{1.65\columnwidth}{!}{
\label{table:regprod1}
\end{table*}

\noindent{\bf The Apostle effect II: Quantifying  the impact of super ties on the long-term citation of individual publications.} 
Determining the impact of super ties on a publication's  long-term   citation tally  is difficult to measure,  because clearly older publications have had more time to accrue citations than newer ones -- a type of censoring bias -- and so a direct comparison of raw citations counts for publications from different years is technically flawed. To address this measurement problem, we map each publication's citation count  ${c}_{i,p,Y}(y)$  in census year $Y_{i}$   to a normalized $z$-score,
\begin{equation}
z_{i,p,y} \equiv  \frac{\ln {c}_{i,p,Y}(y) -   \langle \ln c^{m}_{Y}(y) \rangle}{ \sigma[ \ln c^{m}_{Y}(y)]} \ .
\label{zn}
 \end{equation}
This citation  measure is well-suited for the comparison of publications from different $y$ because $z_{i,p,y}$ is measured relative to the  mean $\langle \ln c^{m}_{Y}(y) \rangle$ number of citations by publications from  the same year $y$, in units of the standard deviation, $\sigma[ \ln c^{m}_{Y}(y)]$  \cite{petersen_inequality_2014}. Thus, we take advantage of the fact that the distribution of citations  obeys a universal log-normal distribution  for $p$ from the same $y$ and discipline  \cite{radicchi_universality_2008}. In this way, $z$ is  defined such that the distribution $P(z)$ is sufficiently  time invariant. To confirm this property, we aggregated $z_{i,p,y}$ within  successive 8-year periods, and calculated the conditional distributions $P(z|y)$, which are stable and  approximately normally distributed over the entire sample period  ({\it SI Text} Fig. S5).

 To define the detrending indices  $\langle ...\rangle$ and $\sigma[...]$ we use the baseline journal set $m$ comprising all research articles collected from the journals  {\it Nature}, {\it Proceedings of the National Academy of Science}, and {\it Science}. We use this aggregation of  three multidisciplinary journals only to control for the time dependent feature of citation counts.
 We chose these journals as our baseline because they have relatively large impact factors (high citation rates), and so the temporal information contained in $\langle ...\rangle$ and $\sigma[...]$ is less noisy than other  $m$ with  lower citation rates. 
Furthermore, since most publications reach their peak citation rate within 5-10 years after publication   \cite{petersen_reputation_2014}, we only analyze $z_{i,p,y}$ with  $y\leq 2003$. In this way, the   $z_{i,p,y}$ values we analyze are less sensitive to fluctuations  early in the citation lifecycle, in addition to recent paradigm shifts in science such as the internet, which  affects the search, the retrieval, and  the citation of prior literature, and the rise of open-access publishing. 
%As such, this  $z$-score   citation measure  is more suitable for intertemporal comparison, because despite the fact that two  publication from different years $y$ and $y'>y$ may have different raw citation counts, as long as both have $z_{p,y}=z_{p',y'}=3$ standard deviations, then they both represent extremely high-impact publications.
 % ranking in the 97.725th percentile of $z$, since $z \approx Normal(0,1)$ distributed variable.} 
%We use the convention of replacing $c_{p}$ by 1 for publications with zero citations; similarly, the mean $\langle \ln c^{ \ j}_{y} \rangle$ and standard deviation $\sigma[ \ln c^{ \ j}_{y}]$ within each journal set are also calculated excluding publications with no citations.  This method of dealing with the logarithm of zero has a negligible overall effect, since only 1.5\% of publications over the time period 1970-2002 had 0 citations in the census year 2009 for the Nat./PNAS/Sci. journal set, and publications in the economics dataset had only twice this frequency. 

In our regression model we use 5 explanatory variables which are author ($i$) and publication ($p$) specific. The first is the number  of coauthors, $a_{i,p}$,  which controls for the tendency for publications with more coauthors to receive more citations \cite{wuchty_increasing_2007}. This variable is also a gross level of technology and coordination costs, since larger teams typically reflect endeavors with higher technical challenge distributed across a wider range of skill sets. We use $\ln a_{i,p}$ since the range of values is rather broad, appearing to be approximately log-normally distributed in the right tail \cite{petersen_quantitative_2014}. 
The second explanatory variable is the dummy variable $R_{i,p}$ which takes the value 1 if  $p$ includes a super tie and the value 0 otherwise. 
Remarkably, the percentage of publications including a super tie is rather close to parity for three of the four datasets: 54\% (top biology), 45\% (top physics), 74\% (other biology) and  54\% (other physics).
The third age variable $t_{i,p}$ is the career age of $i$ at the time of publication.   The fourth variable $N_{i}(t_{p})$  is the total number of publications up to year $t_{i,p}$ which is a non-citation-based measure of  the central author's reputation, visibility, and experience within the scientific community. The final explanatory variable is the collaboration radius,  $S_{i}(t_{p})$, which is the cumulative number of distinct coauthors up to  $t_{i,p}$, representing the central author's access to collaborative resources, as well as an  estimate of the number of researchers in the local community who, having published with $i$, may preferentially cite $i$. Hence, by including $N_{i}(t_{p})$ and  $S_{i}(t_{p})$, we control for two dimensions of   cumulative advantage that could potentially affect a publication's   citation tally.

We then implement a fixed-effects regression to estimate the parameters of the citation impact model, %xtreg zc lkp robinp  taup lcn lck, vce(robust) fe
\begin{eqnarray}
z_{i,p} &=& \beta_{i,0} + \beta_{a} \ln a_{i,p} + \beta_{R} R_{i,p} + \beta_{t} t_{i,p} +  \nonumber \\
& &  \beta_{N} \ln N_{i}(t_{p}) +\beta_{S} \ln S_{i}(t_{p}) + \epsilon_{i,p} \ ,
\label{Reg2}
\end{eqnarray} 
using  the Huber/White/sandwich method to calculate robust standard error estimates that  account for  heteroskedasticity and within-panel serial correlation in the idiosyncratic error term $\epsilon_{i,p}$. %Table \ref{table:regcite} 
We excluded  publications with $y_{p}>2003$,  and in order  that the `top' and `other' datasets are well-balanced, we also excluded the `other' researchers with less than 43 (bio) and 33 (phys.) publications (observations) as of 2003.  Table 1 lists the (standardized) parameter estimates.

%the  ``xtreg , vce(robust) fe'' function in
% STATA11 for each dataset.
%to quantify the effect of super ties on the long-term citation impact of individual publications. 
%This fixed-effects model accounts for the unobserved heterogeneity in time-independent variables related to each researcher profile, assuming that the systemic citation processes are the same for all researchers. 
We estimated  $\beta_{R} = 0.20\pm 0.02 $ ($p\leq0.026$ level in each regression), indicating  a significant  relative citation increase when a publication is  coauthored with at least one super tie.  The standardized  $\beta_{a}$ and  $\beta_{R}$ coefficients are roughly equal, meaning that increasing $a_{p}$ from 1 (a solo author publication) to $e \approx 3$  coauthors produces roughly the same effect as a change in $R_{p}$ from 0 to 1. Thus, while   larger  team size correlates with more citations \cite{wuchty_increasing_2007}, the relative strength of $\beta_{R}$ stresses the importance of `who' in addition to `how many'.

Interestingly, the career age parameter $\beta_{t}=-0.061 \pm 0.004$ is negative  (significant at the $p\leq0.04$ level in each regression), meaning that researchers' normalized citation impact decreases across the career, possibly due to finite career and knowledge life-cycles. This finding is  consistent with a large-scale analysis of  researcher histories within high-impact journals, which also shows  a negative trend in the citation  impact across the career \cite{petersen_inequality_2014}. 
 Neither the reputation ($\beta_{N}$) nor  collaboration radius ($\beta_{S}$) parameters were consistently statistically significant in explaining $z_{i,p,y}$, likely because they are highly correlated with $t_{p}$ for established researchers.
Modifications to consider in followup analysis are controls for the impact factor of the journal publishing $p$, the absolute year $y$ in order to account for shifts in citation patterns in the post-internet era, and removing self-citations from  super ties. Unfortunately, this last task  requires a substantial increase in  data coverage, far beyond the relatively small amount needed to construct  individual ego-network collaboration profiles.
% NOTE: Impact factor may have more of an explanatory importance for domains like economics where an individual publication is considered an opus whereas in the natural sciences individual publications can be rapid, short, incremental, yet still have high impact nevertheless

 We develop three additional descriptive methods in the {\it SI Text} to compare the subset of publications with at least one super-tie  to the complementary subset of publications without one.
These  investigations provide further evidence for the apostle effect.
First, we defined an aggregate career measure, the productivity premium $p_{N,i}$ (see {\it SI Text} Eq.~[S1]),  which measures  the average $K_{ij}$ value among the super ties relative to all the other collaborators. Second,  we defined a similar career measure, the citation premium $p_{C,i}$ (see {\it SI Text} Eq.~[S5]), which quantifies the average citation impact attributable to super ties relative to all the other collaborators.

Independent of dataset, we observed rather substantial  premium values. For example, the  productivity premium has an average value $\langle p_{N} \rangle \approx 8$, meaning that on  a per-collaborator basis, productivity with super ties is roughly 8 times higher than the remaining collaborators. Similarly, the citation premium $p_{C,i}$ is also  significantly right-skewed, with average value $\langle p_{C} \rangle \approx 14$, meaning that net  citation impact per  super tie is 14 times larger than the net  citation impact from all other collaborators. We emphasize that $p_{C,i}$ appropriately accounts for team size by using an equal partitioning of citation credit across the $a_{p}$ coauthors, remedying the multiplicity problem concerning citation credit. 

And third,  we calculated an additional estimation of   the  publication-level citation advantage due to  super ties. For both biology and physics, we found that the publications with super ties  receive roughly 17\% more citations than their counterparts. In basic terms, this means  that the average  publication with a super tie has  21 more citations in biology and  8 more citations in physics than the average publication without  a super tie. This is not a tail effect, because the  citation boost   factor $\alpha_{R}=1.17$ applies a multiplicative shift to the entire citation distribution, $P(\tilde{c} \vert R_{p}=1)\approx P(\alpha_{R}\tilde{c} \vert R_{p}=0)$, thereby impacting  publications above and below the average. \\

\section*{Discussion}
The characteristic  collaboration size in  science has been steadily increasing over the last century \cite{wuchty_increasing_2007,milojevic_principles_2014,petersen_quantitative_2014} with  consequences  at every level of science, from education and academic  careers to universities and funding bodies \cite{pavlidis_together_2014}. 
Understanding how this team-oriented paradigm shift affects the sustainability of careers, the efficiency of the science system, and society's capacity to overcome grand challenges, will be of great importance to a broad range of scientific actors, from scientists to science policy makers.

Collaborative activities are also fundamental to the career growth process, especially in  disciplines where research activities require a division of labor. 
This is especially true in  biology and physics research, where computational, theoretical, and experimental methods provide complementary approaches to a wide array of problems. As a result, a contemporary research group leader is likely to find the assembly of team -- one which is composed of individuals with   diverse yet complementary  skill sets -- a daunting task, especially when under constraints to optimize financial resources,  valuable facilities, and other material resources. Online social network platforms, such as {\it VIVO} (http://www.vivoweb.org/) and {\it Profiles RNS} (http://profiles.catalyst.harvard.edu/), which serve as  match-making recommendation systems, have been developed to facilitate the challenges of team assembly.  

Our analysis indicates that 2/3 of the collaborations analyzed here  are  ``weak''.
Nevertheless, the remaining strong  ties represent social capital investments that can indeed have important  long-term implications, for example  on  information spreading \cite{pan_strength_2012},  career paths \cite{clauset_systematic_2015}, and access to key strategic resources \cite{duch_possible_2012}. 
In  the private sector strong  ties facilitate  access to new growth opportunities, playing an important role in sustaining the competitiveness of firms and employees \cite{uzzi_embeddedness_1999}.  
These considerations further identify why it is important for researchers to understand  the opportunities  that exist within their local network. Understanding the redundancies in the local network \cite{burt_structural_1992} and the   interaction capacity of team members \cite{pentland_new_2012} can help a group leader optimize group intelligence \cite{woolley_evidence_2010} and monitor team efficiency \cite{petersen_persistence_2012}, thereby constituting a source of strategic competitive advantage.  

In summary, we developed methods to better understand the diversity of collaboration strengths. We focused on the career as the unit of analysis, operationalized by using an `ego' perspective so that collaborations, publications, and impact scores fit together into a temporal framework ideal for  cross-sectional and longitudinal modeling.  Analyzing more than 166,000 collaborations, we found that a remarkable   60\%-80\% of the collaborations last only $L_{ij}=1$ year.  Within the subset of repeat collaborations ($L_{ij}\geq $ 2 years), we find that roughly 2/3 of these collaborations last less than a scientist's average duration $\langle L_{i} \rangle \approx $  5 years, yet  1\% last more than $4\langle L_{i} \rangle \approx 20$  years. This wide range in duration and the disparate frequencies of long and short $L_{ij}$, together  point to the dichotomy of burstiness and persistence in scientific collaboration. 
Closer inspection of   individual career paths signals how  idiosyncratic events, such as changing institutions or publishing a seminal study or book, can have significant downstream impact on the arrival rate of new collaboration opportunities and tie formation   (see Figs. \ref{GeimSchematic} and S1).  
Also, the frequency of relatively large publication overlap measures ($f_{K,i}$ and $f_{N,i}$)   indicates that career partners occur rather frequently in science. 

%We observe similar patterns for the economist D. Acemoglu (see Fig. S1), suggesting that our findings are not  merely features of  traditionally collaborative domains, such as biology and physics, but also the social sciences, where reputation, mentoring, collocation, and other features of collaboration are also very important.
In the first part of the study we provide descriptive  insights into basic questions such as how long are typical collaborations, how often does a scientist pair up with his/her main collaborator, and what is the characteristic half-life of a collaboration. 
We also found that as the career progresses, researchers become attractors rather than pursuers of new collaborations. This attractive potential can contribute to cumulative advantage \cite{petersen_quantitative_2011,petersen_inequality_2014}, as it provides select researchers access to a large source of  collaborators, which can boost productivity and increase the potential for a big discovery.

We operationalized  tie strength using an ego-centric perspective of the collaboration network.   
Because the number of publications $K_{ij}$ between the central scientist $i$ and a given coauthor $j$ was found to be   exponentially distributed, the mean value $\langle K_{i} \rangle$ is a natural  author-specific threshold that distinguishes the strong ($K_{ij} \leq \langle K_{i} \rangle$) from the weak ties ($K_{ij} < \langle K_{i} \rangle$). 
Within the subset of strong ties  we  identified   `super tie' outliers  using  an analytic extreme-statistics threshold  $K^{c}_{i}$ defined in  Eq.~[\ref{Kc}].
Also, because the number of publications produced by a collaboration  is highly correlated with its duration, a super tie also represents  persistence that is in excess of the stochastic churn rate that is characteristic of the scientific system.
On a per-collaborator basis, the fraction of coauthors within a research profile that are super ties ($f_{R,i}$) was remarkably common across datasets,  indicating that super ties  occur at an average rate of 1 in 25 collaborators. 
%Because $K^{c}_{i}$ depends only on the distribution of $K_{ij}$ for  $i$,  the super-tie distinction is not symmetric, as it is possible that $j$ is a super tie of $i$, but not vice versa. 

There are various candidate explanations for why such extremely strong collaborations exist. Prosocial motivators may play a strong role, i.e. for some researchers doing science in close community may be more rewarding  than going alone. Also, the search and  formation of a compatible partnership requires time and other social capital investment, i.e. networking. Hence, for two researchers who have found a collaboration that leverages their complementarity, the potential benefits of improving on their match are likely outweighed by the  long-term returns associated with their stable partnership. Complementarity, and the greater skill-set the partnership brings, can also provide a competitive advantage by way of research agility, whereby a larger collective resource base can facilitate rapid adjustments to new and changing knowledge fronts, thereby balancing the risks associated with  changing research direction. After all, a first-mover advantage can make a significant difference in a winner-takes-all credit \& reward system \cite{stephan_how_2012}. 

Scientists may also strategically  pair up in order to share costs, rewards, and risk across the career.
In this light, an additional  incentive to form  super ties may be explained, in part, by the benefits of reward-sharing in the current scientific credit system, wherein publication and  citation credit arising from a single publication are multiplied across the $a_{p}$ coauthors in everyday  practice. Considered in this way, the career risk associated with productivity lulls can  be reduced if a close partnership is formed. 
For example, we observed a few  `twin profiles'  characterized by a publication overlap fraction $f_{K,i}$ between the researcher and his/her  top collaborator that was nearly 100\%. Moreover, we found that 9\% of the biologists and 20\% of the physicists  shared 50\% or more of their papers with their top collaborator.
This highlights a particularly difficult  challenge for science, which  is to develop a credit system which appropriately  divides the net credit, but at the same does not reduce the incentives for scientists to collaborate \cite{pavlidis_together_2014,stallings_determining_2013,allen_credit_2014,shen_collective_2014}.
Thus, it will be important to consider these relatively high levels of publication and citation overlap in the development of  quantitative career evaluation measures, otherwise  there is no penalty to discourage coauthor free-riding \cite{petersen_quantitative_2014}. 

We concluded the analysis by implementing two fixed-effects regression models to determine the  sign and strength of the `apostle effect'  represented by  $\beta_{\rho}$ (productivity) and $\beta_{R}$ (citations). Together, these two coefficients address the fundamental question: is there a measurable advantage associated with heavily investing in a select group of  research partners?

%Motivated by the  descriptive analysis of $L_{ij}$ and $K_{ij}$, we defined multiple longitudinal  variables that quantify a researcher's collaboration   profile. Using these variables as controls, we then implemented two fixed-effects regression models in order to measure the impact of super ties on productivity and citation impact, two dimensions that have become central to  the career evaluation process. The  sign and strength of the `apostle effect' coefficients,  $\beta_{\rho}$ (productivity) and $\beta_{R}$ (impact), which  address the fundamental question: is there any advantage to be gained by heavily investing in a select group of  research life partners?
In the first model we measured the impact of super ties on a researcher's annual publication rate, controlling for career age, average team size,  the prior  experience of $i$ with his/her coauthors, and the relative contribution of super ties within year $t$ as measured by $\rho_{i,t}$ in Eq.~[\ref{rhoeq}].    We found larger $\rho_{i,t}$ to be associated with above-average productivity ($\beta_{\rho}>0$), indicating that super ties play a crucial role in  sustaining  career growth. We also found increased levels of prior experience to be associated with decreased productivity ($\beta_{\overline{L}}<0$), suggesting that maintaining  redundant ties conflicts with the potential benefits from mixing new collaborators into the environment. Nevertheless, higher inequality in the concentration of prior experience was found to have a positive effect on productivity  ($\beta_{G}>0$).

In the second regression model we analyzed the impact of super ties on the citation impact of individual publications, using the detrended citation measure $z_{i,p,y}$ defined in Eq.~[\ref{zn}].   This citation measure is normalized within publication year cohorts, thus allowing for a comparison  of  citation counts  for research articles published in different years. %For each publication $p$, we control for the  team size  $a_{i,p}$, the career year $t_{p}$, and whether or not $p$ was coauthored by at least one  super tie, using the indicator value $R_{p}=1$ in such a case.
%We also included the total number of publications, $N_{i}(t_{p})$, which measures author visibility and reputation, and the collaboration radius (the number of distinct coauthors), $S_{i}(t_{p})$, up to the given year $t_{p}$, which serve as  two cumulative prestige variables.
We found that publications coauthored with super ties, corresponding to  52\%  of the papers we analyzed,  have a significant increase in their long-term citations  ($\beta_{R}>0$). In the {\it SI Text} we  provide additional evidence for the apostle effect, showing  that publications with super ties receive 17\% more citations. This added value may arise from  the extra visibility the publications receives, since the super-tie collaborator may also contribute a substantial reputation and future  productivity that  promote the visibility of the publication. This type of network-mediated reputation spillover is corroborated by a recent study finding a significant citation boost attributable to a researcher's  centrality within the collaboration network   \cite{sarigl_predicting_2014}.   \\
 
 \noindent{\bf Policy recommendations.}
 In all, these results provide quantitative insights into the benefits  associated with strong collaborative partnerships and the value of skill-set complementarity, social trust, and long-term commitment.  This data-oriented analysis also contributes to the  literature on the science of science policy \cite{fealing_science_2011}, providing insight and guidance in an  increasingly metrics-based evaluation system   on how to account for individual achievement in team settings.
One particularly relevant  scenario is fellowship, tenure, and career award evaluations, where it is a common practice  to consider ``independence from one's thesis advisor'' as a selection criteria. We show that in order to assess a researcher's independence,  evaluation committees should also take into consideration 
 the level of publication overlap   between a researcher and his/her strongest collaborator(s). e.g. $f_{K,i}$ and $f_{N,i}$. Yet at the same time, the beneficial role of super ties -- as we have quantitatively demonstrated  -- should also be acknowledged and supported. For example,  funding programs might consider career awards that are specifically multipolar \cite{pavlidis_together_2014}, which would also benefit  the research partners in academia who are  actually life partners, and who may face the daunting ``two-body problem'' of coordinating two research careers. 
Furthermore, understanding the basic levels of publication overlap in science is also important for the ex post facto review of funding outcomes as a means to evaluate the efficiency of science. In large-team settings, measuring the efficiency of a laboratory or project is difficult without a better understanding of how to measure overlapping labor inputs (i.e., collaborator contributions) relative to the project outputs (e.g., publications, patents, etc.). Finally, our study informs early career researchersÑwho are likely to face important decisions concerning the (possibly strategic) selection of collaborative opportunitiesÑon the positive impact that the right research partner can have on their careerÕs long-term sustainability and growth. In all, our results provide quantitative insights into the benefits associated with strong collaborative partnerships, pointing to the added value derived from skill-set complementarity, social trust, and long-term commitment.\\

\noindent {\bf Acknowledgments} The author is grateful for helpful discussions with O. Doria, M. Imbruno, B. Tuncay, and R. Metulini and constructive criticism and keen insights from two anonymous referees. The author also  acknowledges support 
% from the IMT Lucca Foundation and
from the Italian Ministry of Education for the National Research Project (PNR) ``Crisis Lab'' \href{http://www.crisislab.it/}{(http://www.crisislab.it/)} and for feedback from participants  of the  European Union COST Action TD1210 (KnowEscape) workshop on ``Quantifying scientific impact: networks, measures, insights?''

%\bibliography{biblio}
%\small{  % FOR MAKING REFS SMALLER
%\bibliography{MyLibraryOct2014}

\nocite{*}
\begin{thebibliography}{10}

\bibitem{borner_multi-level_2010}
B{\"o}rner K, {et~al.}
\newblock (2010) A multi-level systems perspective for the science of team
  science.
\newblock \emph{Science Translational Medicine} 2:49cm24.

\bibitem{stephan_how_2012}
Stephan P
\newblock (2012) \emph{How Economics Shapes Science}
\newblock (Harvard University Press, Cambridge {MA}, {USA}).

\bibitem{nahapiet_social_1998}
Nahapiet J, Ghoshal S
\newblock (1998) Social capital, intellectual capital, and the organizational
  advantage.
\newblock \emph{Acad. of Management Rev.} 23:242--266.

\bibitem{wuchty_increasing_2007}
Wuchty S, Jones BF, Uzzi B
\newblock (2007) The increasing dominance of teams in production of knowledge.
\newblock \emph{Science} 316:1036--1039.

\bibitem{petersen_reputation_2014}
Petersen AM, {et~al.}
\newblock (2014) Reputation and impact in academic careers.
\newblock \emph{Proceedings of the National Academy of Sciences}
  111:15316--15321.

\bibitem{malmgren_role_2010}
Malmgren RD, Ottino JM, Amaral LAN
\newblock (2010) The role of mentorship in protege performance.
\newblock \emph{Nature} 463:622--626.

\bibitem{petersen_quantitative_2014}
Petersen AM, Pavlidis I, Semendeferi I
\newblock (2014) A quantitative perspective on ethics in large team science.
\newblock \emph{Sci. \& Eng. Ethics.} 20:923--945.

\bibitem{pavlidis_together_2014}
Pavlidis I, Petersen AM, Semendeferi I
\newblock (2014) Together we stand.
\newblock \emph{Nature Physics} 10:700--702.

\bibitem{borgatti_network_2009}
Borgatti SP, Mehra A, Brass DJ, Labianca G
\newblock (2009) Network analysis in the social sciences.
\newblock \emph{Science} 323:892--895.

\bibitem{granovetter_strength_1973}
Granovetter MS
\newblock (1973) The strength of weak ties.
\newblock \emph{Amer. J. Sociology} 78:1360--Ñ1380.

\bibitem{newman_structure_2001}
Newman MEJ
\newblock (2001) The structure of scientific collaboration networks.
\newblock \emph{Proceedings of the National Academy of Sciences} 98:404--409.

\bibitem{newman_scientific_2001-3}
Newman MEJ
\newblock (2001) Scientific collaboration networks. {I}. network construction
  and fundamental results.
\newblock \emph{Phys. Rev. E} 64:016131.

\bibitem{Barabasi_evolution_2002}
Barabasi AL, {et~al.}
\newblock (2002) Evolution of the social network of scientific collaborations.
\newblock \emph{Physica A: Statistical Mechanics and its Applications} 311:590
  -- 614.

\bibitem{newman_coauthorship_2004}
Newman MEJ
\newblock (2004) Coauthorship networks and patterns of scientific
  collaboration.
\newblock \emph{Proceedings of the National Academy of Sciences}
  101:5200--5205.

\bibitem{guimera_team_2005}
Guimera R, Uzzi B, Spiro J, Amaral LAN
\newblock (2005) Team assembly mechanisms determine collaboration network
  structure and team performance.
\newblock \emph{Science} 308:697--702.

\bibitem{palla_quantifying_2007}
Palla G, Barabasi AL, Viscek T
\newblock (2007) Quantifying social group evolution.
\newblock \emph{Nature} 446:664--667.

\bibitem{pan_strength_2012}
Pan RK, Saram\"aki J
\newblock (2012) The strength of strong ties in scientific collaboration
  networks.
\newblock \emph{{EPL}} 97:18007.

\bibitem{martin_coauthorship_2013}
Martin T, Ball B, Karrer B, Newman MEJ
\newblock (2013) Coauthorship and citation patterns in the physical review.
\newblock \emph{Phys. Rev. E} 88:012814.

\bibitem{ke_tie_2014}
Ke Q, Ahn YY
\newblock (2014) Tie strength distribution in scientific collaboration
  networks.
\newblock \emph{Phys. Rev. E} 90:032804.

\bibitem{borner_simultaneous_2004}
B{\"o}rner K, Maru JT, Goldstone RL
\newblock (2004) The simultaneous evolution of author and paper networks.
\newblock \emph{Proceedings of the National Academy of Sciences}
  101:5266--5273.

\bibitem{milojevic_principles_2014}
Milojevic S
\newblock (2014) Principles of scientific research team formation and
  evolution.
\newblock \emph{Proceedings of the National Academy of Sciences}
  111:3984--3989.

\bibitem{march_exploration_1991}
March JG
\newblock (1991) Exploration and exploitation in organizational learning.
\newblock \emph{Organizational Science} 2:71--87.

\bibitem{lazer_network_2007}
Lazer D, Friedman A
\newblock (2007) The network structure of exploration and exploitation.
\newblock \emph{Adm. Sci. Quarterly} 52:667--694.

\bibitem{petersen_persistence_2012}
Petersen AM, Riccaboni M, Stanley HE, Pammolli F
\newblock (2012) Persistence and uncertainty in the academic career.
\newblock \emph{Proc. Natl. Acad. Sci. {USA}} 109:5213 -- 5218.

\bibitem{pentland_new_2012}
Pentland A
\newblock (2012) The new science of building great teams.
\newblock \emph{Harvard Business Review} 90:60--69.

\bibitem{woolley_evidence_2010}
Woolley AW, et al.
\newblock (2010) Evidence for a collective intelligence factor in the
  performance of human groups.
\newblock \emph{Science} 330:686--688.

\bibitem{stallings_determining_2013}
Stallings J, {et~al.}
\newblock (2013) Determining scientific impact using a collaboration index.
\newblock \emph{Proceedings of the National Academy of Sciences}
  110:9680--9685.

\bibitem{allen_credit_2014}
Allen L, Brand A, Scott J, Altman M, Hlava M
\newblock (2014) Credit where credit is due.
\newblock \emph{Nature} 508:312--313.

\bibitem{shen_collective_2014}
Shen HW, Barabasi AL
\newblock (2014) Collective credit allocation in science.
\newblock \emph{Proceedings of the National Academy of Sciences}
  111:12325--12330.

\bibitem{petersen_quantitative_2011}
Petersen AM, Jung WS, Yang JS, Stanley HE
\newblock (2011) Quantitative and empirical demonstration of the {M}atthew effect
  in a study of career longevity.
\newblock \emph{Proceedings of the National Academy of Sciences} 108:18--23.

\bibitem{petersen_inequality_2014}
Petersen AM, Penner O
\newblock (2014) Inequality and cumulative advantage in science careers: a case
  study of high-impact journals.
\newblock \emph{{EPJ} Data Science} 3:24.

\bibitem{krapivsky_kinetic_2010}
Krapivsky P, Redner S, Ben-Naim E
\newblock (2010) \emph{A kinetic view of statistical physics}
\newblock (Cambridge University Press, Cambridge, {UK}).

\bibitem{azoulay_superstar_2010}
Azoulay P, Zivin JSG, Wang J
\newblock (2010) Superstar extinction.
\newblock \emph{Q. J. of Econ.} 125:549--589.

\bibitem{Uzzi_Atypical_2013}
Uzzi B, Mukherjee S, Stringer M, Jones B
\newblock (2013) Atypical combinations and scientific impact.
\newblock \emph{Science} 342:468--472.

\bibitem{radicchi_universality_2008}
Radicchi F, Fortunato S, Castellano C
\newblock (2008) Universality of citation distributions: Toward an objective
  measure of scientific impact.
\newblock \emph{Proc. Natl. Acad. Sci. {USA}} 105:17268--17272.

\bibitem{clauset_systematic_2015}
Clauset A, Arbesman S, Larremore DB
\newblock (2015) Systematic inequality and hierarchy in faculty hiring
  networks.
\newblock \emph{Science Advances} 1.

\bibitem{duch_possible_2012}
Duch J, {et~al.}
\newblock (2012) The possible role of resource requirements and academic
  career-choice risk on gender differences in publication rate and impact.
\newblock \emph{{PLoS} One} 125:e51332.

\bibitem{uzzi_embeddedness_1999}
Uzzi B
\newblock (1999) Embeddedness in the making of financial capital: How social
  relations and networks benefit firms seeking financing.
\newblock \emph{Amer. Soc. Rev.} 64:481--505.

\bibitem{burt_structural_1992}
Burt RS
\newblock (1992) \emph{Structural Holes}
\newblock (Harvard University Press, Cambridge {MA}, {USA}).

\bibitem{sarigl_predicting_2014}
Sarigl E, Pfitzner R, Scholtes I, Garas A, Schweitzer F
\newblock (2014) Predicting scientific success based on coauthorship networks.
\newblock \emph{{EPJ} Data Science} 3:9.

\bibitem{fealing_science_2011}
Fealing KH, {eds.}
\newblock (2011) \emph{The science of science policy: A handbook.}
\newblock (Stanford Business Books, Stanford {CA}, {USA}).

% SI ONLY BELOW: 42-
\begin{comment}
\bibitem{petersen_distribution_2008}
Petersen AM, Jung WS, Stanley HE
\newblock (2008) On the distribution of career longevity and the evolution of
  home run prowess in professional baseball.
\newblock \emph{{EPL}} 83:50010.


\bibitem{petersen_methods_2011}
Petersen AM, Penner O, Stanley HE
\newblock (2011) Methods for detrending success metrics to account for
  inflationary and deflationary factors.
\newblock \emph{Eur. Phys. J. B} 79:67--78.


\bibitem{petersen_methods_2010}
Petersen AM, Wang F, Stanley HE
\newblock (2010) Methods for measuring the citations and productivity of
  scientists across time and discipline.
\newblock \emph{Phys. Rev. E} 81:036114.

\bibitem{petersen_statistical_2011}
Petersen AM, Stanley HE, Succi S
\newblock (2011) Statistical regularities in the rank-citation profile of
  scientists.
\newblock \emph{Scientific Reports} 1:181.

\bibitem{petersen_z-index:_2013}
Petersen AM, Succi S
\newblock (2013) The {Z}-index: A geometric representation of productivity and
  impact which accounts for information in the entire rank-citation profile.
\newblock \emph{Journal of Informetrics} 7:823 -- 832.

\bibitem{Penner_predictability_2013}
Penner O, Pan RK, Petersen AM, Fortunato S
\newblock (2013) On the predictability of future impact in science.
\newblock \emph{Scientific Reports} 3:3052.


\bibitem{acemoglu_economic_2005}
Acemoglu D, Robinson JA
\newblock (2005) \emph{Economic Origins of Dictatorship and Democracy}
\newblock (Cambridge University Press, Cambridge, {UK}).

\end{comment}
\end{thebibliography}
%}
%\bibliographystyle{naturemag}
\bibliographystyle{pnas}

\end{document}